\documentclass[a4paper, 11pt]{article}
\usepackage{graphicx}
\usepackage{bm}
\usepackage{amsmath, amssymb, amsthm}
\usepackage{subfig}
\usepackage{color}
\usepackage{mathrsfs}
\usepackage{multirow}

\newcommand\bx{\boldsymbol{x}}

\newcommand\bu{\boldsymbol{u}}
\newcommand\bv{\boldsymbol{v}}

\newcommand\bE{\boldsymbol{E}}
\newcommand\bj{\boldsymbol{j}}
\newcommand\bB{\boldsymbol{B}}
\newcommand\bxi{\boldsymbol{\xi}}

\newcommand\bbR{\mathbb{R}}
\newcommand\bbN{\mathbb{N}}

\newcommand\bq{\boldsymbol q}

\newcommand\dd{\,\mathrm{d}}

\newcommand\He{\mathit{He}}

\newcommand\mO{\mathcal{O}}
\newcommand\mM{\mathcal{M}}

\newcommand\balpha{{\alpha}}

\newcommand\mfq{\mathfrak{q}}

\newcommand{\imag}{\mathrm{i}}

\newcommand{\mH}{\mathcal{H}}
\newcommand{\mT}{\mathcal{T}}
\newcommand\pd[2]{\dfrac{\partial {#1}}{\partial {#2}}}

\newcommand\od[2]{\dfrac{\dd {#1}}{\dd {#2}}}
\newcommand\bsigma{{\boldsymbol \sigma}}
\newcommand\bI{\boldsymbol{I}}
\numberwithin{equation}{section}

\graphicspath{{images/}} 
{\theoremstyle{remark} \newtheorem{remark}{Remark}}

\title{A Two-fluid Model for Plasma with Prandtl Number Correction}

\author{Ruo Li\thanks{CAPT, LMAM \& School of Mathematical Sciences,
    Peking University, Beijing, China, email: {\tt
      rli@math.pku.edu.cn}.}, ~~Yixiao Lu\thanks{School of
    Mathematical Sciences, Peking University, Beijing, China, 100871,
    email: {\tt luyixiao@pku.edu.cn}.},~~ Yanli Wang\thanks{Beijing
    Computational Science Research Center, email: {\tt
      ylwang@csrc.ac.cn}.} }

\begin{document}
\maketitle
\begin{abstract}

  A two-fluid model is derived from the plasma kinetic equations using
  the moment model reduction method. The moment method we adopt was
  recently developed with a globally hyperbolic regularization where
  the moment model attained is locally well-posed in time. Based on
  the hyperbolic moment model with well-posedness, the Maxwellian
  iteration method is utilized to get the closure relations for the
  resulted two-fluid model. By taking the Shakhov collision operator
  in the Maxwellian iteration, the two-fluid model inherits the
  correct Prandtl number from the plasma kinetic equations. The new
  model is formally the same as the five-moment two-fluid model except
  for the closure relations, where the pressure tensor is anisotropic
  and the heat flux is presented. This provides the model the capacity
  to depict problems with anisotropic pressure tensor and large heat
  flux.

  \vspace*{4mm}
  \noindent {\bf Keywords:} Plasma kinetic equations; Maxwellian
  iteration; two-fluid model; moment model reduction

\end{abstract}

\section{Introduction}

The multi-species plasma model was proposed to capture dynamics in
systems with multiple species of charges \cite{Krall1973}. Each
species is treated separately and coupled together through
electromagnetic and collisional processes. In the collisionless
problems, the evolution may be most accurately \cite{Petkaki2006,
  Vann2003} modeled using the kinetic theory such as the
Vlasov-Maxwell equations, where the plasma is described by the
distribution functions of all species, and the electromagnetic fields
are governed by Maxwell equations. However, it is extremely expensive
to numerically solve the plasma kinetic equations due to the high
dimensionality of the plasma kinetic equations\cite{SHUMLAK2003}.

Several reduced models were introduced \cite{Hakim2008, Miller2016,
  HAKIM2006} to approximate the plasma kinetic equations. When plasma
behaves like an electrically conducting fluid, where the motion of the
electrons and ions is locked together by electrostatic forces, the MHD
model is utilized \cite{Friedberg1987}. In the MHD model, the plasma
is treated as a single electrically conducting fluid. Several
algorithms have been designed based on the MHD model, which are
already used to simulate several plasma phenomena successfully
\cite{implicitJones1997, transport1998}. It was pointed out that the
MHD model ignores the electron mass and the finite Larmor radius
effects \cite{kineticCheng1999}. This may lead to the limitation to
treat the Hall effect and diamagnetic terms \cite{Tilt2000}. Hall
effect can be added to the MHD equations and we get the Hall-MHD model
\cite{Huba2003}. Though a distinction is made between the bulk plasma
velocity and electron velocity in the Hall-MHD model, electron inertia
and displacement current are still ignored, and the plasma is assumed
to be quasi-neutral \cite{HAKIM2006}.

In order to capture the separate motion of the electrons and ions, the
two-fluid plasma model was proposed without adding the complexity of
the kinetic model \cite{SHUMLAK2003}. The electromagnetic fields are
also modeled using Maxwell equations of electromagnetism. This model
was derived \cite{Mason1986, Mason1987} by taking moments of the
plasma kinetic equations for each species and is also a generalization
of the MHD model. The two-fluid plasma model retains electron inertia
effects and displacement current. However, directly taking moments of
the plasma kinetic equations will lead to an unclosed equation system
and different closure methods will deduce various kinds of two-fluid
plasma models. For example, assuming there is no heat flow, one
obtains the five-moment ideal two-fluid equations with scalar fluid
pressure\cite{HAKIM2006, SHUMLAK2003}. The strongly collisional plasma
can be accurately described by the five-moment model. In the weaker
collisional regime, the anisotropies increase that the five-moment
model is not valid any more \cite{HAKIM2006}. Several higher-order
moment descriptions of the plasma have been developed respectively for
the collisionless and collisional transport in plasma. For example,
the ten- and sixteen- moment hydrodynamic models are derived for the
collisionless regime \cite{Comparision2015, Landau1997,
  Oraevskii_1968}. The ten- and thirteen- moment models for capturing
collisional transport in mixed gases and magnetic plasma are also
derived \cite{Miller2016, Hakim2008, tippett_1995, Johnson2011,
  Gilliam2011}.

For reduced models involving only a few macroscopic variables, it is
of great importance to predict the physical Prandtl number. However,
only a few works have been done to preserve the correct Prandtl
number, though various versions of the two-fluid model have been
derived. In this paper, we are aiming at deriving a new two-fluid
model based on the plasma kinetic model with the correct Prandtl
number. The method we adopt is based on the moment model reduction
method developed in \cite{Fan_new, Wang} and the classical Maxwellian
iteration \cite{Ikenberry, Truesdell}. First, the distribution
function is expanded into a Hermite series \cite{Wang}, and the
hyperbolic moment equations (HME) are derived for the plasma kinetic
model \cite{VM2015} under the framework \cite{framework}. The method
in \cite{framework} was first developed for the Boltzmann equation,
and the distribution function is approximated by the Hermite expansion
around local Maxwellian. The globally hyperbolic regularization method
\cite{Fan_new} is adopted to get the hyperbolic moment equations,
which is locally well-posed in time. Here we follow the same approach
therein to get the HME system for the plasma kinetic model with
Shakhov collision operator. Then, the Maxwellian iteration is applied
to get the closure relations for the shear stress and heat flux based
on the HME system. These closure relations have the same form as that
in \cite{Callen2005}, but the Prandtl number in the Shakhov collision
model is inherited to the resulted two-fluid model. Consequently, the
new model is equipped with the capacity to predict the correct Prandtl
number, which is also formally validated by the demonstrative
examples. In comparison, this new reduced model is formally the same
as the five-moment model in \cite{HAKIM2006} except for the closure
relations. The five-moment model has isotropic pressure and neglects
the heat flux. In this new model, the pressure tensor is anisotropic
and the heat flux is presented, where both terms are expressed by the
density, macroscopic velocity and temperature. We hope that the new
reduced model may provide improved performance for the problem with
not negligible anisotropic pressure tensor and heat flux for different
Prandtl numbers.

The rest of this paper is arranged as follows. In Section
\ref{sec:pre}, the two-fluid model for the plasma and the plasma
kinetic model are introduced briefly. In Section \ref{sec:expansion},
we make the Hermite expansion to the kinetic plasma model and then
derive the hyperbolic moment equations (HME). The new reduced
two-fluid model is deduced in Section \ref{sec:closure}, where the
closure relations for the shear stress and the heat flux are derived
by Maxwellian iteration. We present demonstrative examples in Section
\ref{sec:numerical} to validate our new model. A short conclusion in
Section \ref{sec:conclusion} closes the paper.


\section{Preliminary} \label{sec:pre}

In the collisionless plasma where the collective interactions dominate
the plasma dynamics, the Vlasov-Maxwell (VM) equations are mostly used
to describe the evolution of the plasma. When the distribution
function of the particles is not far from Maxwellian, the movement of
the plasma can be described by the two-fluid model \cite{HAKIM2006,
  fullLaguna2018}. In this section, we will introduce the two-fluid
plasma model and the VM model briefly.

\subsection{Two-fluid plasma model}
\label{sec:two-fluid}
The two-fluid model, which captures the separate motion of the
electrons and ions, is derived by taking moments of the plasma kinetic
model for each species. This process eliminates the microscopic
velocity space, and at the same time, the macroscopic variables,
including the density, momentum and energy, are utilized to describe
the evolution of the plasma. The detailed form of the two-fluid model
is as below
\begin{equation}
  \label{eq:two_fluid}
  \begin{aligned}
 &   {\rm density:} & & \od{n_{\beta}}{t}  + n_{\beta}(\nabla_{\bx} \cdot
    \bu_{\beta}) = 0,\\
 &   {\rm momentum:} & & m_{\beta} n_{\beta} \od{\bu_{\beta}}{t} =
    n_{\beta} \mfq_{\beta} [\bE + \bu_{\beta} \times \bB] -
    \nabla_{\bx} p_{\beta} - \nabla_{\bx}\cdot \bsigma_{\beta}, \\
 &   {\rm energy:}& & \frac{3}{2}n_{\beta} \od{\mT_{\beta}}{t} = -
    n_{\beta} \mT_{\beta} (\nabla_{\bx} \cdot  \bu_{\beta})-
    \nabla_{\bx} \cdot \bq_{\beta} - \bsigma_{\beta} \cdot \nabla_{\bx}
    \bu_{\beta},
  \end{aligned}
\end{equation}
where $\beta = i$ or $e$, represents electrons and ions
respectively. The expression $\bsigma \cdot \nabla_{\bx} \bu$ is
defined as
\begin{equation}
  \label{eq:sigma_u}
  \bsigma \cdot \nabla_{\bx} \bu =
  \sum_{i,j=1}^3\sigma_{ij} \pd{u_j}{x_i}.
\end{equation}
$m_{\beta}$ and $\mfq_{\beta}$ is the mass and charge of the
particles. $n_{\beta}$, $\bu_{\beta}$ and $\mT_{\beta}$ represent the
density, macroscopic velocity and temperature. Moreover, $p_{\beta}$
is the isotropic pressure, $\bsigma_{\beta}$ is the shear stress
tensor, and $\bq_{\beta}$ is the heat flux. $\bE$ and $\bB$ are the
electric and magnetic field respectively, which is decided by the
Maxwell equations, the exact form of which is as below
\begin{equation}
  \label{eq:Maxwell_1}
  \left\{
    \begin{array}{c}
      \pd{\bB}{t} + \nabla_{\bx} \times \bE = 0, \\[4mm]
      \dfrac{1}{c^2}\pd{\bE}{t} - \nabla_{\bx} \times \bB = -\nu_0 \bj, \\[4mm]
      \nabla_{\bx} \cdot \bE = \dfrac{\rho_c}{\epsilon_0}, \qquad
      \nabla_{\bx} \cdot \bB = 0.
    \end{array}
  \right.
\end{equation}
Here $\nu_0$ and $\epsilon_0$ are the permeability and permittivity of
free space \cite{Hakim2008}, and $c = (\nu_0\epsilon_0)^{-1/2}$ is the
speed of light.  $\rho_c$ and $\bj$ are the charged density and
current density defined by
\begin{equation}
  \label{eq:current_1}
  \rho_c = \sum_{\beta =i, e}\mfq_{\beta}n_{\beta}, \qquad
  \bj(t, \bx)  =\sum_{\beta =i, e}\mfq_{\beta} n_{\beta}\bu_{\beta}.
\end{equation}

The two-fluid model \eqref{eq:two_fluid} is not closed, and there are
several kinds of closure methods, which will lead to different
two-fluid models. For example, in the five-moment ideal two-fluid
model, the anisotropic stress $\bsigma_{\beta}$ and the heat flux
$\bq_{\beta}$ are set as zero. Combined with the equation of state for
the ideal gas, we can get the closed five-moment ideal two-fluid model
\cite{HAKIM2006}. Similarly, we can also get the ten-moment two-fluid
equations \cite{Hakim2008}. In \cite{Callen2005}, the closure of
$\bsigma$ and $\bq$ are derived in the kinetic theory of gas with the
BGK collision term
\begin{equation}
  \label{eq:closure}
  \bq = -\kappa \nabla_{\bx} \mT, \qquad \bsigma = -2 \mu \left(\frac{1}{2}
  \left[\nabla_{\bx} \bu + (\nabla_{\bx} \bu)^T\right] -\frac{1}{3} \bI
  (\nabla_{\bx} \cdot \bu)\right),
\end{equation}
where $\kappa$ is the coefficient of heat conductivity and $\mu$ is
the coefficient of viscosity. Since only one species is considered in
the kinetic theory of gas, the footnote index $\beta$ in
\eqref{eq:closure} is eliminated. However, in \eqref{eq:closure} the
BGK collision model will lead to wrong Prandtl number, which equals
$1$ for the ideal gas, while the correct one equals $2/3$. Therefore,
other collision models should be utilized to get the correct Prandtl
number.

Based on the Shakhov collision model, we will try to deduce the
two-fluid model with the correct Prandtl number based on the kinetic
plasma equations. In the next section, the plasma kinetic equations
will be introduced briefly.

\subsection{Plasma kinetic equations}
In the plasma kinetic equations, the plasma is described by the
distribution functions $f(t, \bx, \bv)$, which depends on time,
physical space and microscopic velocity space. The plasma kinetic
equations then have the form below
\begin{equation}
  \label{eq:kinetic}
  \pd{f_{\beta}}{t} + \bv \cdot \nabla_{\bx} f_\beta
  + \frac{\mfq_\beta}{m_\beta}(\bE + \bv \times \bB) \cdot
  \nabla_{\bv} f_\beta
  = \pd{f_{\beta}}{t}\Big\vert_{\rm collision}, \qquad (\bx, \bv) \in
  \Omega \times \bbR^3,
\end{equation}
where $\beta$ also represents the ions, or electrons respectively, and
$\Omega \in \bbR^3$. The electron-magnetic fields $(\bE, \bB)$ are
given by the classical Maxwell system \eqref{eq:Maxwell_1}.  In the VM
model, where for the low-collisional plasma, the collisions are
neglected and collective interactions are assumed to dominate the
plasma dynamics, which means that
\begin{equation}
  \label{eq:col}
  \pd{f_{\beta}}{t}\Big\vert_{\rm collision} = 0.
\end{equation}
In what follows, we are focusing on the derivation of the closed
two-fluid model with correct Prandtl number. Therefore, the Shakhov
collision model is applied \cite{Shakhov}, whose exact form is
\begin{equation}
  \label{eq:Shakhov}
  Q_{\rm Shakhov}(f_{\beta}) = \nu_{\beta}(f^{s}_{\beta} - f_{\beta}), \qquad
  f^{s}_{\beta} = P_{\beta}(t, \bx, \bv)\mM_{\beta}(t, \bx, \bv),
\end{equation}
with
\begin{equation}
  \label{eq:P3}
  P_{\beta} = 1 + \frac{(1 - {\rm Pr})(\bv - \bu_{\beta}) \cdot
    \bq_{\beta}}{(D+2) \rho_{\beta}(t, \bx)\mT_{\beta}(t,
    \bx)^2}\left(\frac{|\bv -\bu_{\beta}|^2}{\mT_{\beta}(t, \bx)} -
    (D + 2)\right), \qquad D = 3,
\end{equation}
where $\nu_{\beta}$ is the collision frequencies,
$D$ is the dimension number of the microscopic velocity space
and $\mM_{\beta}(t, \bx, \bv)$ is Maxwellian, which has the form
\begin{equation}
  \label{eq:maxwellian_mul}
  \mM_{\beta}(t, \bx, \bv) = \frac{n_{\beta}}{(2\pi
    \mT_{\beta})^{3/2}}\exp\left(-\frac{(\bv - \bu_{\beta})^2}{2\mT_{\beta}} \right).
\end{equation}
Moreover, $\rm Pr$ is the Prandtl number which is decided by the type
of the particles. $\bu_{\beta}$, $\mT_{\beta}$ and $\bq_{\beta}$ are
the macroscopic variables defined in the last section, whose
relationships with the distribution function $f_{\beta}(t, \bx, \bv)$
are as below
\begin{equation}
  \label{eq:macro_f}
  \begin{gathered}
    n_{\beta} = \int_{\bbR^3} \bv f_{\beta} \dd \bv, \qquad n_{\beta}
    \bu_{\beta} = \int_{\bbR^3} \bv f_{\beta} \dd \bv, \qquad
    \frac{3}{2}n_{\beta} \mT_{\beta}
    = \frac{1}{2} \int_{\bbR^3} |\bv - \bu_{\beta}|^2 f_{\beta} \dd \bv, \\
    \bq_{\beta}= \frac{1}{2}\int_{\bbR^3}|\bv - \bu|^2(\bv -
    \bu_{\beta}) f_{\beta} \dd \bv, \qquad p_{\beta, ij} =
    \int_{\bbR^3} (v_i - u_{\beta,i}) (v_j - u_{\beta,j}) f_{\beta} \dd \bv, \qquad
    i,j=1,2,3.
  \end{gathered}
\end{equation}
The pressure $p_{\beta} $ is defined as
\begin{equation}
  \label{eq:pressure}
  p_{\beta}= \frac{1}{3}\sum_{i=1}^{3} p_{\beta, ii}.
\end{equation}

Since the plasma kinetic equation is seven-dimensional, it is only
used to simulate the plasma to capture the essential physics
\cite{SHUMLAK2003}. Some simplified models are introduced to simulate
the evolution of the plasma, just by taking moments of the plasma
kinetic equations.

Different collision models and closure methods will lead to different
two-fluid models. In the following sections, we will consider the
Shakhov collision model purposely for a correct Prandtl number in the
reduced model.


\section{Moment Model Reduction}
\label{sec:expansion}
In this section, we will first derive the hyperbolic moment model for
the plasma kinetic equations, which is the base for us to deduce the
new two-fluid model. For simplicity, we consider at first the
single-species case of the non-relativistic electrons under the
self-consistent electromagnetic fields while the ions are treated as a
uniform fixed background. The whole procedure can be trivially
extended to the general plasma.

Without loss of generality, we consider the dimensionless form of the
governing equations for one species. Eliminating the footnote index
for the distribution function $f_{\beta}(t, \bx, \bv)$, the
dimensionless equation for the electrons is as below
\begin{equation}
  \label{eq:VM}
  \pd{f}{t} + \bv \cdot \nabla_{\bx} f + (\bE + \bv \times \bB) \cdot
  \nabla_{\bv} f = \nu(f^s- f), \qquad (\bx,  \bv)
  \in  \Omega \times \bbR^3,
\end{equation}
and the dimensionless form for the classical Maxwell system is
\begin{equation}
  \label{eq:Maxwell}
  \left\{
    \begin{array}{c}
      \pd{\bE}{t} - \nabla_{\bx} \times \bB = -\bj, \\[4mm]
      \pd{\bB}{t} + \nabla_{\bx} \times \bE = 0, \\[4mm]
      \nabla_{\bx} \cdot \bE = \rho - h, \qquad \nabla_{\bx} \cdot \bB = 0,
    \end{array}
  \right.
\end{equation}
where the relations of the current density $\bj$ and charge density
$\rho$ with the distribution function $f$ are reduced into
\begin{equation}
  \label{eq:current}
  \begin{aligned}
    \bj(t, \bx)  &= \int_{\bbR^3} f(t, \bx, \bv) \bv \dd \bv, \qquad
  \rho(t, \bx)  = \int_{\bbR^3} f(t , \bx, \bv) \dd \bv,
\end{aligned}
\end{equation}
with $h$ the initial background density satisfying
\begin{equation}
  \label{eq:back}
  \int_{\Omega} (\rho - h) \dd \bx = 0.
\end{equation}
The relationship between the macroscopic variables and the
distribution function is listed in \eqref{eq:macro_f}.  Besides, the
Maxwellian is reduced into
\begin{equation}
  \label{eq:maxwellian}
  \mM(t, \bx, \bv) = \frac{\rho}{(2\pi
    \mT)^{3/2}}\exp\left(-\frac{(\bv - \bu)^2}{2\mT} \right).
\end{equation}

Following the method in \cite{Wang}, we approximate the distribution
function using Hermite basis functions as
\begin{equation}
  \label{eq:expansion}
  f(\bv) \approx \sum_{|\alpha| \leqslant
    M}f_{\alpha}\mathcal{H}_{\mT,\alpha}(\bxi), \qquad
  \bxi = \frac{\bv-\bu}{\sqrt{\mT}},
\end{equation}
where $M$ is the expansion order, and
$\alpha=(\alpha_1,\alpha_2,\alpha_3)$ is a three-dimensional
multi-index. The basis functions $\mathcal{H}_{\mT,\alpha}$ are
defined as
\begin{equation}
  \label{eq:base}
  \mathcal{H}_{\mT,\alpha}(\bxi) =\prod\limits_{d=1}^3
 \frac{1}{\sqrt{2\pi}}\mT^{-\frac{\alpha_d+1}{2}}
\He_{\alpha_d}(\xi_d)\exp \left(-\frac{\xi_d^2}{2} \right),
\end{equation}
where $\He_{\alpha_d}$ is the Hermite polynomial
\begin{equation}
  \label{eq:hermit}
  \He_n(x) = (-1)^n\exp \left( \frac{x^2}{2} \right) \frac{\dd^n}{\dd
    x^n} \exp \left(-\frac{x^2}{2} \right).
\end{equation}
For convenience, $\He_{n}(x)$ is taken as zero if $n<0$, thus
$\mathcal{H}_{\mT,\alpha}(\bxi)$ is zero when any component of
$\alpha$ is negative.  Based on the expansion \eqref{eq:expansion}, we
can find that the Maxwellian $\mM$ in \eqref{eq:maxwellian} equals
\begin{equation}
  \label{eq:maxwellian_term}
  \mM(t, \bx, \bv) = f_0(t, \bx)\mathcal{H}_{\mT,0}(\bxi), \qquad   \bxi =
  \frac{\bv-\bu}{\sqrt{\mT}},
\end{equation}
and it holds for the coefficients $f_{\alpha}$ that
\begin{equation}
\label{eq:estimate}
  f_0 = \rho(t,\bx), \quad f_{e_i} = 0, \quad
  \sum\limits_{d=1}^3 f_{2e_d} = 0, \quad i =1, 2 , 3.
\end{equation}
Moreover, the relationship between heat flux $q_i$, shear stress
$\sigma_{ij}$ and the expansion coefficients $f_{\alpha}$ are
\begin{gather}
  \label{eq:q_p}
  \sigma_{ij} = (1 + \delta_{ij}) f_{e_i + e_j}, \qquad q_{i} =
  2f_{3e_i} + \sum\limits_{d=1}^3f_{2e_d+e_i}, \qquad i,j=1,2,3,
\end{gather}
where the relationship between the shear stress $\sigma_{ij}$ and the
distribution function is
\begin{equation}
  \label{eq:stress}
  \sigma_{ij} = \int_{\bbR} \left((v_i - u_i)(v_j - u_j) -
  \frac{1}{3}\delta_{ij}|\bv - \bu|^2\right) f \dd \bv, \qquad i,j=1,2,3.
\end{equation}
With the equation of state of the ideal gas, we can derive the
relationship between the pressure tensor and the shear stress as
\begin{equation}
  \label{eq:pressure_eos}
  p = \rho \mT, \qquad   \sigma_{ij} = p_{ij} - \delta_{ij}p.
\end{equation}

Below we briefly introduce the moment systems for \eqref{eq:VM}, and
refer \cite{VM2015} for the detailed procedure. With the relationship
\begin{gather}
  \label{eq:basis_function_relation}
  \pd{}{v_j}\mH_{\mT,\balpha}\left(\bxi\right)
  = -\mH_{\mT, \balpha+e_j}\left( \bxi \right),  \\
  v_j \mH_{\mT,\balpha}\left(\bxi\right) = \mT \mH_{\mT,
    \balpha+e_j}\left(\bxi\right) + u_j\mH_{\mT,
    \balpha}\left(\bxi\right) + \balpha_{j}\mH_{\mT,
    \balpha-e_j}\left(\bxi\right),
\end{gather}
we can get that
\begin{align}
  \label{eq:EB_relation}
  & \nabla_{\bv}\cdot \big [(\bE + \bv \times \bB)f\big] = \\ \nonumber
  & \quad - \sum_{d=1}^3
    \left[E_d - \sum_{k,m}^3\epsilon_{dkm}u_kB_m\right]f_{\balpha - e_d}
    - \sum_{d, k,m=1}^3 \epsilon_{dkm}(\alpha_k + 1)B_m f_{\balpha - e_d
    + e_k},
\end{align}
where the Levi-Civita symbols $\epsilon_{dkm}$ are defined as
\begin{equation}
  \label{eq:Levi-Civita}
  \epsilon_{dkm} = \left\{
    \begin{array}{rl}
      1, & d \neq k \neq m ~\text{cyclic permutation of } 1, 2, 3, \\
      -1, & d \neq k \neq m ~\text{anti-cyclic permutation of } 1, 2,
            3, \\
      0, & (d-k)(k - m)(m -d) = 0.
    \end{array}\right.
\end{equation}
The Shakhov collision term can be expanded as
\begin{equation}
  \label{eq:collision}
  \nu(f^s - f) =  \nu \sum_{\alpha \in
    \bbN^3} Q_{\alpha}\mathcal{H}_{\mT,\alpha}(\bxi), \qquad   \bxi =
  \frac{\bv-\bu}{\sqrt{\mT}},
\end{equation}
with
\begin{equation}
  \label{eq:collision_coe}
  Q_{\alpha} = \left\{
    \begin{array}{cc}
      0, & |\alpha| <2, \\
      \dfrac{1-{\rm Pr}}{5} q_j -f_{\alpha},
         &\alpha = 2e_i + e_j, \quad i, j = 1, 2, 3, \\
      -f_{\alpha}, & {\rm otherwise}.
    \end{array}
 \right.
\end{equation}
By plugging the expansion \eqref{eq:expansion} into \eqref{eq:VM}, the
general moment equations can be obtained as
\begin{equation}
  \label{eq:moment_system_inf}
  \begin{split}
     \pd{f_{\balpha}}{t} &+ \sum_{d =
      1}^3\left[\pd{u_d}{t}+\sum_{j=1}^3u_j\pd{u_d}{x_j} - E_d -
      \sum_{k,m=1}^3\epsilon_{dkm}u_kB_m\right] f_{\balpha-e_d} \\
    & - \sum_{d,k,m=1}^3 \epsilon_{dkm}(\alpha_k + 1) B_m
    f_{\balpha-e_d + e_k} + \frac{1}{2}\left(\pd{\mT}{t} +
      \sum_{j=1}^3u_j\pd{\mT}{x_j}\right)\sum_{d=1}^3
    f_{\balpha-2e_d} \\
    &  + \sum_{j,d=1}^3\left[\pd{u_d}{x_j}\left(\mT
        f_{\balpha-e_d-e_j}
        + (\alpha_j + 1)f_{\balpha-e_d + e_j}(1 - H[|\alpha| - M])\right) \right. \\
    &  \left. + \frac{1}{2}\pd{\mT}{x_j}\left(\mT f_{\balpha -
          2e_d -e_j} +
        (\alpha_j + 1)f_{\balpha - 2e_d + e_j}(1 - H[|\alpha| - M])\right)\right]  \\
    &  + \sum_{j = 1}^3 \left(\mT\pd{f_{\balpha-e_j}}{x_j} + u_j
      \pd{f_{\balpha}}{x_j} + (\alpha_j +
      1)\pd{f_{\balpha+e_j}}{x_j}(1 - H[|\alpha| - M])\right) =
    \nu Q_{\alpha},\quad 0 \leqslant |\balpha|\leqslant M,
  \end{split}
\end{equation}
where $M$ is the expansion order and $H[x]$ is defined as
\begin{equation}
  \label{eq:Hx}
  H[x] = \left\{
    \begin{array}{cc}
      1, & x = 0, \\
      0, & {\rm otherwise}.
    \end{array}
\right.
\end{equation}
Similar to that in \cite{Wang}, we can deduce the conservation of
mass, momentum and the equation for the temperature as
\begin{equation}
  \label{eq:mass_conservation}
  \begin{gathered}
  \pd{\rho}{t} + \sum_{j=1}^3 \left(u_j \pd{\rho}{x_j} + \rho
    \pd{u_j}{x_j}\right) = 0,\\
  \pd{u_d}{t} + \sum_{j=1}^3 u_j \pd{u_d}{x_j} + \frac{1}{\rho}\sum_{j
    = 1}^3 \pd{p_{jd}}{x_j} = E_d + \sum_{k,m}^3\epsilon_{dkm}u_kB_m,
  \\
  \rho\left(\pd{\mT}{t} + \sum_{j=1}^3u_j\pd{\mT}{x_j}\right) +
  \frac{2}{3}\sum_{j=1}^3 \left(\pd{q_j}{x_j} + \sum_{d = 1}^3
    p_{jd}\pd{u_d}{x_j} \right) = 0.
  \end{gathered}
\end{equation}
Substituting $p_{ij}$ in \eqref{eq:mass_conservation} with
\eqref{eq:stress}, we can derive that
\begin{equation}
  \begin{gathered}
    \label{eq:vel_T_final}
    \pd{\rho}{t} + \sum_{j=1}^3 \left(u_j \pd{\rho}{x_j} + \rho
      \pd{u_j}{x_j}\right) = 0,\\
    \pd{u_d}{t} + \sum_{j=1}^3 u_j \pd{u_d}{x_j} +
    \frac{1}{\rho}\sum_{j = 1}^3 \left(\pd{\sigma_{jd}}{x_j} +
      \delta_{jd}\pd{p}{x_j}\right) = E_d +
    \sum_{k,m}^3\epsilon_{dkm}u_kB_m,
    \\
    \rho\left(\pd{\mT}{t} + \sum_{j=1}^3u_j\pd{\mT}{x_j}\right) +
    \frac{2}{3}\sum_{j=1}^3 \left(\pd{q_j}{x_j} + \sum_{d = 1}^3
      \left(\sigma_{jd}+\delta_{jd} p \right)\pd{u_d}{x_j} \right) =
    0.
  \end{gathered}
\end{equation}
Finally, substituting $\pd{u_d}{t}$ and $\pd{\mT}{t}$ in
\eqref{eq:moment_system_inf} with \eqref{eq:vel_T_final}, we can get
the quasi-linear system for the higher order moments of the Shakhov
collision term as
\begin{equation}
  \label{eq:moment_system}
  \begin{split}
    F_{\alpha} &= -\nu f_{\alpha}, \qquad \forall ~|\alpha| = 2,\\
    F_{\alpha} &= \nu \left(\frac{1-\Pr}{5}q_j - f_{\alpha}\right),
    \qquad \alpha =
    2e_i + e_j, \quad i, j = 1,2,3,\\
    F_{\alpha} &= -\nu f_{\alpha}, \qquad \forall ~3 \leqslant
    |\alpha| \leqslant M~{\rm and}~\alpha \neq 2e_i + e_j, \quad i, j
    = 1, 2,3,
\end{split}
\end{equation}
where
\begin{equation}
  \label{eq:left}
  \begin{split}
& F_{\alpha} = \pd{f_{\alpha}}{t}+\sum_{j=1}^3\left(
  \mT\pd{f_{\alpha-e_j}}{x_j}+u_j\pd{f_{\alpha}}{x_j}
  +(\alpha_j+1)\pd{f_{\alpha+e_j}}{x_j}(1 - H[|\alpha| - M])\right) \\
&\qquad +\sum_{j=1}^3\sum_{d=1}^3\left( \mT f_{\alpha-e_d-e_j} +
  (\alpha_j+1)f_{\alpha-e_d+e_j}(1 - H[|\alpha| - M])
  -\frac{p_{jd}}{3\rho}\sum_{k=1}^3f_{\alpha-2e_k}\right)\pd{u_d}{x_j}\\
&\qquad+\sum_{j=1}^3\left(\sum_{k=1}^3\big(\mT
    f_{\alpha-2e_k-e_j}+(\alpha_j+1)f_{\alpha-2e_k+e_j}(1 - H[|\alpha| - M])\big)
  \left(-\frac{\mT}{2\rho}\pd{\rho}{x_j}
    +\frac{1}{6\rho}\sum_{d=1}^3 \pd{p_{dd}}{x_j}\right)\right)\\
&\qquad -\sum_{j=1}^3\sum_{d=1}^3
\frac{f_{\alpha-e_d}}{\rho}\pd{p_{jd}}{x_j}
-\frac{1}{3\rho}\left(\sum_{k=1}^3f_{\alpha-2e_k}\right)\sum_{j=1}^3\pd{q_j}{x_j}
- \sum_{d,k,m=1}^3 \varepsilon_{dkm} \left(\alpha_k +
  1\right)B_m f_{\balpha - e_d + e_k},
\end{split}
\end{equation}

Together with \eqref{eq:vel_T_final} and \eqref{eq:moment_system}, we
derive the hyperbolic moment system for the one species plasma kinetic
equations \eqref{eq:VM}.

It is known that one can deduce the Euler and Navier-Stokes equations
from the Boltzmann equation based on the zeroth- and first-order
expansions of the distribution function. For the similarity of the
Boltzmann and the plasma kinetic equation, we will try to deduce the
two-fluid plasma model based on the moment equations.

\section{New Two-Fluid Model}
\label{sec:closure}
To derive our new two-fluid model based on HME for the plasma kinetic
equations, we carry out a Maxwellian iteration to give the closure
relations. The Maxwellian iteration was introduced by Ikenberry and
Truesdell \cite{Ikenberry, Truesdell} as a technique to derive the NSF
and Burnett equations from the moment equations. It is then used to
analyze the order for the magnitude of each moment, which is known as
the COET (Consistently Ordered Extended Thermodynamics) method.

Here, Maxwellian iteration is utilized to derive models with the first
order of accuracy. To begin with, we introduce the scaling
$t = t' / \epsilon$ and $x_i = x_i' / \epsilon$, and rewrite the
moment equations \eqref{eq:moment_system} with time and spatial
variables $t'$ and $x_i'$.  Thus, a factor $\epsilon^{-1}$ is
introduced to the right-side of \eqref{eq:moment_system}. In this
section, we will work on the scaled equations, and the prime symbol
$t'$ and $x'$ will be omitted. With some rearrangement,
\eqref{eq:moment_system} is rewritten as
\begin{equation}
  \label{eq:moment_system_BGK}
  \begin{split}
    F_{\alpha} &= -\frac{\nu}{\epsilon} f_{\alpha}, \qquad \forall
    ~|\alpha| = 2, \\
    F_{\alpha} &= \frac{\nu}{\epsilon} \left(\frac{1-\Pr}{5}q_j -
      f_{\alpha}\right), \qquad \alpha =
    2e_i + e_j, \quad i, j = 1,2,3, \\
    F_{\alpha} &= -\frac{\nu}{\epsilon} f_{\alpha}, \qquad \forall ~3
    \leqslant |\alpha| \leqslant M,~{\rm and}~\alpha \neq 2e_i + e_j,
    \quad i, j = 1,2,3.
  \end{split}
\end{equation}
Assuming that the asymptotic expansions for all the moments have the
form
\begin{equation}
  \label{eq:ce_expansion}
  f_{\alpha} = f_{\alpha}^{(0)} + \epsilon f_{\alpha}^{(1)} +
  \epsilon^2 f_{\alpha}^{(2)} + \cdots,
\end{equation}
we will begin the iteration based on \eqref{eq:moment_system_BGK}.  In
the original Maxwellian iteration method, we require that $f^{(0)}$ is
the local Maxwellian $\mM(t, \bx, \bv)$. The corresponding assumption
for the coefficients is
\begin{equation}
  \label{eq:iter_f0}
  f_{\alpha}^{(0)} =  \left\{
    \begin{array}{cc}
      \rho, & {\rm if~} |\alpha| = 0, \\
      0, & {\rm otherwise.}
    \end{array}\right.
\end{equation}
Then noting that $f_{\alpha} = 0$ if $|\alpha| =1$ based on
\eqref{eq:estimate}, the iteration begins from $|\alpha| \geqslant 2$,
which can be written as the iterative scheme below
\begin{equation}
  \label{eq:iteration}
  f_{\alpha}^{(n+1)} = - \mathcal{G}_{\alpha}\left(f_{\beta}^{(n)}
  |\, \beta \in \bbN^3\right), \quad \forall \alpha \in \bbN^3~{\rm and~}
  |\alpha| \geqslant 2, \qquad n = 0, 1, 2, \cdots,
\end{equation}
and the macroscopic variables $\rho$, $\bu$ and $\mT$ are treated as
\begin{equation}
  \label{eq:macro}
  \rho =\mO(1),   \quad \bu =\mO(1), \quad \mT = \mO(1).
\end{equation}
When $|\alpha| =2$,
\eqref{eq:moment_system_BGK}  is changed into
\begin{equation}
  \label{eq:iter_f2_1}
  \begin{split}
    f_{\alpha} = -\frac{\epsilon}{\nu} F_{\alpha} , \qquad \forall
    ~|\alpha| = 2.
  \end{split}
\end{equation}
In the iteration, $\{f_{\alpha}, |\alpha| > 2\}$ are treated as of
high order, and the terms $\{f_{\alpha}, |\alpha| = 2\}$ in
$F_{\alpha}$ are also omitted, due to the $\epsilon$ on the right hand
side. Thus, we can easily deduce the approximation to $f_{\alpha}$
when $|\alpha| =2$ as
\begin{equation}
  \label{eq:iter_1_f2}
  \begin{aligned}
    f_{2e_i}^{(1)} &= -\frac{1}{\nu} \left[
      -\frac{1}{3}\sum_{j=1}^{3}\left(\sum_{d=1}^3 p_{jd}\pd{u_d}{x_j}
        + \pd{q_j}{x_j} \right) + \rho \mT
      \pd{u_i}{x_i}\right], \qquad i = 1, 2, 3,\\
    f_{e_i + e_k}^{(1)} &= -\frac{1}{\nu}\left[ \mT \rho
      \left(\pd{u_d}{x_j} + \pd{u_j}{x_d} \right)\right], \qquad i, k
    = 1, 2, 3, \quad i \neq k.
  \end{aligned}
\end{equation}
Similarly, when $|\alpha| = 3$, \eqref{eq:moment_system_BGK}  is
changed into
\begin{equation}
  \label{eq:iter_f2_2}
  \begin{split}
    f_{\alpha} &= -\frac{\epsilon}{\nu} F_{\alpha} + \frac{1 -
      \Pr}{5}q_j, \qquad \alpha = 2e_i + e_j, \quad i, j = 1,
    2, 3, \\
    f_{\alpha}& = -\frac{\epsilon}{\nu} F_{\alpha}, \qquad \alpha =
    e_i + e_j + e_k, \quad i, j, k = 1, 2, 3, \quad i \neq j \neq k.
  \end{split}
\end{equation}
From \eqref{eq:q_p}, we can find that $\bq$ is at the same order as
$\{f_{\alpha}, |\alpha| = 3\}$. Assuming that the asymptotic expansion
for heat flux $q_i$ is
\begin{equation}
  \label{eq:ce_expansion_q}
  q_i = q_i^{(0)} + \epsilon q_i^{(1)} +
  \epsilon^2 q_i^{(2)} + \cdots,\quad i = 1, 2, 3,
\end{equation}
then $q_i^{(0)} = 0$ due to \eqref{eq:iter_f0}. Thus it holds for
$\{f_{\alpha}^{(1)}, |\alpha| = 3\}$ that
\begin{equation}
  \label{eq:iter_1_f3_tmp}
  \begin{aligned}
    f_{2e_i + e_k}^{(1)} &= -\frac{1}{\nu}\mT
      \left(-\frac{\mT}{2}\pd{\rho}{x_k} + \frac{1}{6}\sum_{d=1}^3
        \pd{p_{dd}}{x_k} \right)+\frac{1-{\rm Pr}}{5}q_k^{(1)}
,\qquad i, k = 1, 2,3, \\
    f_{e_i + e_j + e_k}^{(1)} &= 0, \qquad i, j, k = 1, 2,3, \quad i
    \neq j \neq k.
\end{aligned}
\end{equation}
With the relationship between the pressure tensor and the temperature
\eqref{eq:pressure} and \eqref{eq:pressure_eos},
\eqref{eq:iter_1_f3_tmp} is reduced into
\begin{equation}
  \label{eq:iter_1_f3}
  \begin{aligned}
    f_{2e_i + e_k}^{(1)}
    &=-\frac{1}{2\nu}\rho\mT\pd{\mT}{x_k}+\frac{1-{\rm Pr}}{5}q_k^{(1)}
     ,\qquad i, k = 1, 2,3, \\
    f_{e_i + e_j + e_k}^{(1)} &= 0, \qquad i, j, k = 1, 2,3, \quad i
    \neq j \neq k.
\end{aligned}
\end{equation}
Moreover, with the same method, we can derive that for the higher
order of coefficients,
\begin{equation}
  \label{eq:iter_1_f4}
  f_{\alpha}^{(1)} = 0, \qquad |\alpha| \geqslant 4.
\end{equation}
For now, the first order of iteration is finished.
From \eqref{eq:q_p} and the expression of $f_{\alpha}^{(1)}$, we can
get the approximation to the shear stress and heat flux at order
$\mathcal{O}(\epsilon)$
\begin{align}
  \label{eq:shear_f1_1}
    \sigma_{ii} &= 2 f_{2e_i} \approx -\frac{2\epsilon}{\nu} \left[
      -\frac{1}{3}\sum_{j=1}^{3}\left(\sum_{d=1}^3 p_{jd}\pd{u_d}{x_j}
        + \pd{q_j}{x_j} \right) + \rho \mT
      \pd{u_i}{x_i}\right], \qquad i = 1, 2, 3,\\
  \label{eq:shear_f1_2}
  \sigma_{ij} &= f_{e_i + e_j} \approx - \frac{\epsilon}{\nu}\rho \mT \left(\pd{u_i}{x_j}
                + \pd{u_j}{x_i} \right), \qquad  i \neq j,  \qquad i, j = 1, 2,
                3, \\
  \label{eq:q_f1}
  q_i & = 2 f_{3e_i} + \sum_{d = 1}^3 f_{2e_d+e_i} \approx -\frac{1}{{\rm Pr}}
    \frac{5}{2}\frac{\epsilon}{\nu}
        \rho \mT \pd{\mT}{x_i}, \qquad i = 1, 2, 3.
\end{align}
From \eqref{eq:shear_f1_2} and \eqref{eq:q_f1}, we can find that
$\sigma_{ij}$ and $q_i$ are all at order
$\mathcal{O}(\epsilon)$. Substituting \eqref{eq:shear_f1_2} and
\eqref{eq:q_f1} into \eqref{eq:shear_f1_1} and omitting the terms at
order $\mO(\epsilon^2)$, we can get the simplified form of
\eqref{eq:shear_f1_1} with \eqref{eq:stress}
\begin{equation}
  \label{eq:shear_f1_1_final}
  \sigma_{ii}  \approx -\frac{2\epsilon}{\nu} \rho \mT \left(\pd{u_i}{x_i} -
    \frac{1}{3}\sum_{j=1}^3\pd{u_j}{x_j}\right), \qquad i = 1, 2, 3.
\end{equation}
Introducing the parameter viscosity conductivity $\mu$ and thermal
conductivity $\kappa$, we can get the final closure term by letting
$\epsilon \rightarrow 1$ as
\begin{equation}
  \label{eq:closure_term}
  \begin{gathered}
    \sigma_{ij} = - \mu \left(\pd{u_i}{x_j} + \pd{u_j}{x_i} -
      \frac{2}{3}\delta_{ij}\sum_{k=1}^3\pd{u_k}{x_k}\right), \qquad
    q_i = -\kappa \pd{\mT}{x_i}, \qquad i = 1, 2, 3,
  \end{gathered}
\end{equation}
where
\begin{equation}
  \label{eq:viscosity}
  \mu = \frac{\rho\mT}{\nu}, \qquad \kappa =
  \frac{1}{\Pr}\frac{5}{2}\frac{\rho\mT}{\nu}.
\end{equation}

\begin{remark}
  From the definition of Prandtl number
  \begin{equation}
    \label{eq:Pr}
    \Pr = \frac{5}{2}\frac{\mu}{\kappa},
  \end{equation}
  we can find that \eqref{eq:viscosity} could give the correct Prandtl
  number for any type of particles. However, if the BGK collision
  model is utilized instead of the Shakhov collision model in the
  deduction, the wrong Prandtl number $\Pr = 1$ will be derived.
\end{remark}

For the moment, we have got the approximation of the macroscopic
variables such as the shear stress $\sigma_{ij}$ and the heat flux
$q_i$ with the density $\rho$, macroscopic velocity $\bu$ and
temperature $\mT$. Next, we will deduce the new reduced model with
these approximations and the moment equations
\eqref{eq:mass_conservation} and \eqref{eq:vel_T_final}.  Rewriting
\eqref{eq:vel_T_final} into the vector form and adopting
\eqref{eq:closure_term}, we can get the equation system as
\begin{equation}
  \label{eq:closed_system}
  \left\{
    \begin{array}{c}
      \pd{\rho}{t} + \nabla_{\bx} \cdot (\rho \bu) = 0, \\
      \rho \od{\bu}{t} = -\nabla_{\bx} p  - \nabla_{\bx} \cdot \bsigma
      + \rho (\bE + \bu \times \bB), \\
      \dfrac{3}{2}\rho \od{\mT}{t} = [(-p \bI - \bsigma) \cdot \nabla_{\bx}]
      \cdot \bu - \nabla_{\bx} \cdot \bq,
    \end{array}
\right.
\end{equation}
where $\od{\cdot}{t} = \pd{\cdot}{t} + \bu \cdot \nabla_{\bx}{\cdot}$,
and $\bI$ is the $3 \times 3$ unit matrix with
\begin{equation}
  \label{eq:closure_HME}
  \begin{gathered}
    \bsigma = (\sigma_{ij})_{3 \times 3} = - \mu \left[\nabla_{\bx}
      \bu + (\nabla_{\bx} \bu)^T -
      \frac{2}{3}(\nabla_{\bx} \cdot \bu)\bI  \right], \\
    p  = \rho \mT, \qquad \bq = -\kappa \nabla_{\bx} \mT,  \qquad
    \mu  = \frac{\rho\mT}{\nu}, \qquad \kappa =
  \frac{1}{\Pr}\frac{5}{2}\frac{\rho\mT}{\nu}.
  \end{gathered}
\end{equation}
Until now, we have deduced the reduced model for
\eqref{eq:VM}. Compared with the two-fluid model in Section
\ref{sec:two-fluid}, we can find \eqref{eq:closure_HME} has the same
form of \eqref{eq:closure}, but with different coefficients of
viscosity and thermal conductivity. This is due to the different
collision models and closure methods, where the Shakhov collision
model is adopted here and will inherit the correct Prandtl number from
the plasma kinetic equations.

Since the shear stress and the heat flux are all expressed by density,
macroscopic velocity and the temperature, this new reduced model has
the same number of degree of freedom as the five-moment two-fluid
model, which greatly decreases the computational complexity compared
with the plasma kinetic equations. More importantly, the shear stress
and heat flux are not simply set as zero, which is done in the
five-moment model \cite{SHUMLAK2003}, but expressed by the macroscopic
variables such as the density, macroscopic velocity and
temperature. Thus, it is expected that this new reduced model could be
more capable to describe the problems with anisotropic pressure
tenser and large heat flux compared to the five-moment model.

Though this new model is only deduced for the single-species of
plasma, it can be extended naturally to the plasma kinetic equation
\eqref{eq:kinetic}. For the collisionless plasma where the collective
interactions dominate the plasma dynamics, we can neglect the momentum
and heat transfer between different species. Thus, we can derive the
new two-fluid model for multi-species of plasma with the same
Maxwellian iteration. In this case, the new reduced model
\eqref{eq:closed_system} is changed into
\begin{equation}
  \label{eq:two_fluid_1}
  \begin{aligned}
    & {\rm density:} & & \od{n_{\beta}}{t} + n_{\beta}(\nabla_{\bx}
    \cdot
    \bu_{\beta}) = 0,\\
    & {\rm momentum:} & & m_{\beta} n_{\beta} \od{\bu_{\beta}}{t} =
    n_{\beta} \mfq_{\beta} [\bE + \bu_{\beta} \times \bB] -
    \nabla_{\bx} p_{\beta} - \nabla_{\bx}\cdot \bsigma_{\beta}, \\
    & {\rm energy:}& & \frac{3}{2}n_{\beta} \od{\mT_{\beta}}{t} = -
    n_{\beta} \mT_{\beta} (\nabla_{\bx} \cdot \bu_{\beta})-
    \nabla_{\bx} \cdot \bq_{\beta} - \bsigma_{\beta} \cdot
    \nabla_{\bx} \bu_{\beta},
  \end{aligned}
\end{equation}
with the closure relations
\begin{equation}
  \label{eq:closure_two-fluid}
  \begin{gathered}
    \bsigma_{\beta} = - \mu_{\beta} \left[\nabla_{\bx} \bu_{\beta} +
      (\nabla_{\bx} \bu_{\beta})^T -
      \frac{2}{3}(\nabla_{\bx} \cdot \bu_{\beta})\bI  \right], \\
    p = n_{\beta} \mT_{\beta}, \qquad \bq_{\beta} = -\kappa_{\beta}
    \nabla_{\bx} \mT_{\beta} , \qquad \mu_{\beta} =
    \frac{n_{\beta}\mT_{\beta}}{\nu_{\beta}}, \qquad \kappa_{\beta} =
    \frac{1}{\Pr}\frac{5}{2}\frac{n_{\beta}\mT_{\beta}}{\nu_{\beta}}.
  \end{gathered}
\end{equation}

\begin{remark}
  We only derive the new reduced model for the plasma which does not
  contain the interactions between different species. For the general
  case, the corresponding new reduced model can also be derived but
  with some differences in the closure relations.

  For the different collision models, the same Maxwellian iteration
  method can also be utilized to get the reduced models for the plasma
  kinetic equations, but the final form of the two-fluid model may be
  different.
\end{remark}


\section{Model Validation}
\label{sec:numerical}
In this section, the validation of this new model is studied,
especially for the problem where Prandtl number will bring negligible
effect. To show the capability of the new model, the steady-state and
time-dependent problems are tested.

\paragraph{First example}
In this example, we focus on the steady-state problem. The Hartmann
flow problem is often used to test the resistive, viscous MHD models
near the high collision regime \cite{implicitJones1997,
  Miller2016}. Below, a similar problem is examined to show that the
new two-fluid model can capture correct Prandtl number.

\begin{figure}[!htb]
  \centering
  \includegraphics[width=0.35\textwidth, height=0.25\textwidth,
    clip]{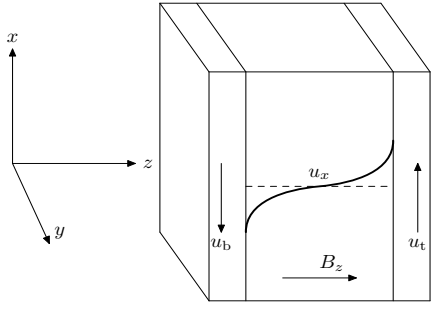}
    \caption{Hartmann problem.}
  \label{fig:Hartmann}
\end{figure}

The problem setting shown in Figure \ref{fig:Hartmann} is a plasma set
between two infinitely large plates. Two infinite conducting plates
are driving shear flow along $x$-direction with different velocities
and temperatures. A constant magnetic field is imposed along
$z$-direction, which drives a current in $y$-direction. The flow
generated along $y$ is interacting with $B_z$, which will suppress the
viscous boundary layer.  Since we are focusing on the effect of
Prandtl number, the dimensionless problem of only one species with a
fixed background is tested. To further reduce this problem, we just
set the viscosity and thermal conductivity $\mu$ and $\kappa$ in
\eqref{eq:two_fluid_1} as constant. Moreover, the walls are no-slip,
but no adiabatic.

For the steady state problem, the governing equations are reduced into
\begin{equation}
  \label{eq:steady}
  \begin{aligned}
    & \sum_{j=1}^3\pd{}{x_j} (\rho u_j)= 0, \\
    & \sum_{j=1}^3u_j \pd{u_d}{x_j} +
    \frac{1}{\rho}\sum_{j=1}^3\left(\pd{\sigma_{jd}}{x_j} + \delta_{jd}
      \pd{p}{x_j}\right) = E_d + \sum_{k,m}^3 \epsilon_{dkm} u_k B_m, \\
    & \rho\sum_{j=1}^3u_j\pd{\mT}{x_j} +
    \frac{2}{3}\sum_{j=1}^3\left(\pd{q_j}{x_j} + \sum_{d=1}^3(\sigma_{jd} +
    \delta_{jd} p) \pd{u_d}{x_j}\right) = 0,
  \end{aligned}
\end{equation}
with the closure
\begin{equation}
  \label{eq:closure_steady}
  \begin{aligned}
    p = \rho \mT, \qquad \bsigma = -\mu \left[\nabla_{\bx} \bu +
    (\nabla_{\bx} \bu)^T - \frac{2}{3} (\nabla_{\bx} \cdot \bu) \bI\right],
    \qquad \bq = -\kappa \nabla_{\bx} \mT,
  \end{aligned}
\end{equation}
where $x_1, x_2, x_3$ are corresponding to $x, y, z$. Different
boundary conditions will lead to different steady state solutions.  In
this test, the plasma is assumed to be incompressible and the initial
density is set as $\rho = 1$. The velocity and temperature of the
upper wall are set as $u_{\rm up} = 1$, and $\mT_{\rm up} = 1$, while
those of the bottom wall are set as $u_{\rm bottom} = 0$, and
$\mT_{\rm bottom} = 0$.  The height between the two plates is $L = 1$,
with the imposed magnetic field $B_z = 1$. For this problem is
designed for a slab geometry, there is no gradient in the $x$ or $y$
direction, which means that
\begin{equation}
  \label{eq:slab}
\pd{\cdot}{x} = \pd{\cdot}{y} = 0.
\end{equation}
Then from the continuity equation, we can derive
\begin{equation}
  \label{eq:uz}
  u_z = 0.
\end{equation}
Steady state Faraday's Law in a slab geometry \cite{Miller2016} could
give
\begin{equation}
  \label{eq:slab_E}
  E_x = E_y = 0.
\end{equation}
Then the governing system \eqref{eq:steady} is reduced into
\begin{equation}
  \label{eq:reduced}
  \begin{aligned}
    &\mu\pd{^2 u_x}{z^2} = -\rho u_y B_z, \\
    &\mu\pd{^2 u_y}{z^2} = \rho u_x B_z, \\
    &\kappa \pd{^2 \mT}{z^2} = -\mu \left( \left(\pd{u_x}{z}\right)^2
      + \left(\pd{u_y}{z}\right)^2\right).
  \end{aligned}
\end{equation}
Then we can get the solution to the steady state problem as
\begin{equation}
  \label{eq:solution_final}
  \begin{aligned}
    u_x(z) & = a_1 \exp\left(bz\right) + a_2\exp\left(-bz\right) + a_3
    \exp\left( \bar{b} z\right) + a_4  \exp\left(-\bar{b} z\right),  \\
    u_y(z) & = -\imag \big[a_1 \exp\left(b z\right) + a_2\exp\left(-b
      z\right) - a_3
    \exp\left( \bar{b} z\right) - a_4  \exp\left(-\bar{b} z\right)\big],  \\
    \mT(z)& =-\frac{4{\rm
        Pr}}{5}a_1a_3\Big[\exp\left((b+\bar{b})z\right)
    -\exp\left((b-\bar{b})z\right) \\
    & \qquad \qquad  \qquad -\exp\left((-b+\bar{b})z\right)
    +\exp\left((-b-\bar{b})z\right) \Big]+Cz,
  \end{aligned}
\end{equation}
where
\begin{equation}
  \begin{split}
      \label{eq:coe}
      a_1 &=\frac{1}{2\left(\exp(b)-\exp(-b)\right)},
      \quad a_2=-\frac{1}{2\left(\exp(b)-\exp(-b)\right)}, \\
      a_3 &=\frac{1}{2\left(\exp(\bar{b})-\exp(-\bar{b})\right)},
      \quad a_4=-\frac{1}{2\left(\exp(\bar{b})-\exp(-\bar{b})\right)},
      \qquad b  = \frac{1}{\sqrt{2 \mu}}(1 + \imag), \\
      C & =1+\frac{4 {\rm
        Pr}}{5}a_1a_3\left[\exp\left(-\frac{\sqrt{2}}{\sqrt{\mu}}\right)
      -\exp\left(-\frac{\sqrt{2} \imag}{\sqrt{\mu}}\right)
      -\exp\left( \frac{\sqrt{2} \imag}{\sqrt{\mu}}\right)
      +\exp\left(\frac{\sqrt{2}}{\sqrt{\mu}}\right)\right].
    \end{split}
  \end{equation}

\begin{figure}[!htb]
  \centering
  \includegraphics[width=0.35\textwidth, height=0.25\textwidth,
    clip]{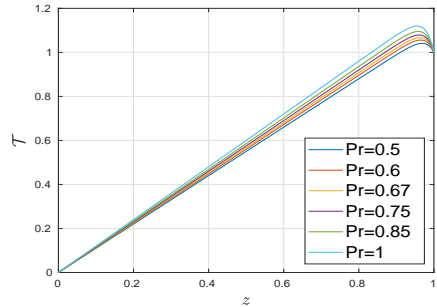}
    \caption{Temperature for different Prandtl numbers.}
    \label{fig:T}
  \end{figure}

  In Figure \ref{fig:T}, the solutions of the temperature for
  different Prandtl numbers are plotted where the viscous conductivity
  is set as $\mu = 0.01$. We can see that the appearance of the
  temperature with different Prandtl numbers varies greatly, which
  means Prandtl number is quite important to get the correct
  model. The new reduced two-fluid model could inherit the correct
  Prandtl number from the plasma kinetic equations, and could be more
  capable to describe the behavior of the plasma.

  \paragraph{Second example}

  \begin{figure}[!htb]
    \centering
    \includegraphics[width=0.35\textwidth, height=0.25\textwidth,
    clip]{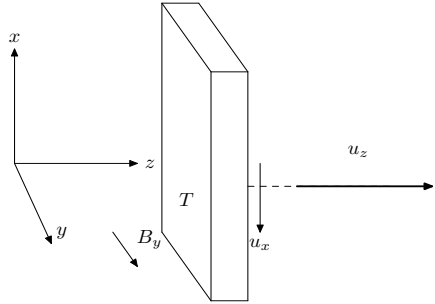}
    \caption{Time-dependent problem.}
    \label{fig:test2}
  \end{figure}

\begin{figure}[!htb]
  \centering
    \subfloat[$\rho$]
  {\includegraphics[width=0.25\textwidth, height=0.2\textwidth,
    clip]{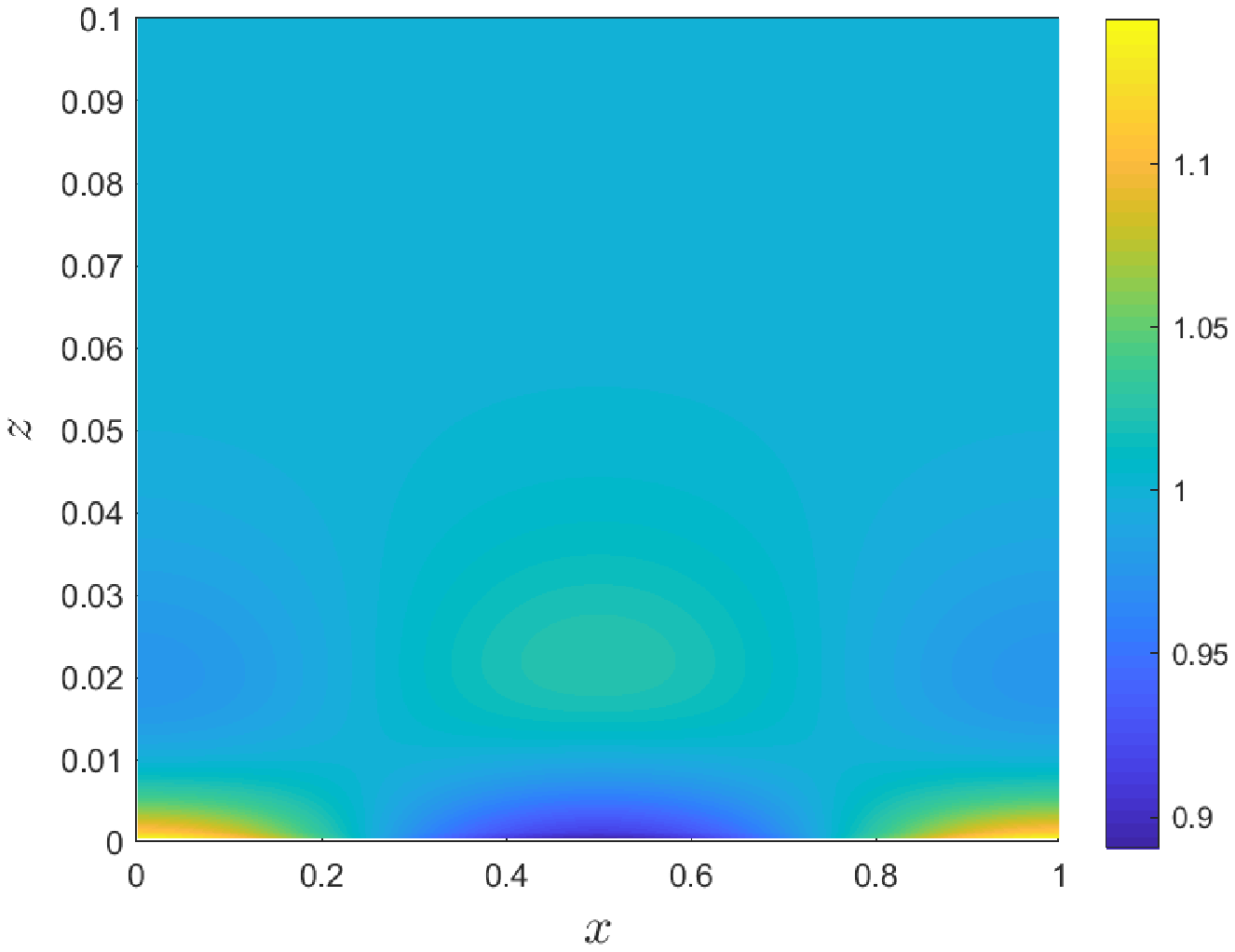}}\hfill
    \subfloat[$u_1$]
  {\includegraphics[width=0.25\textwidth, height=0.2\textwidth,
    clip]{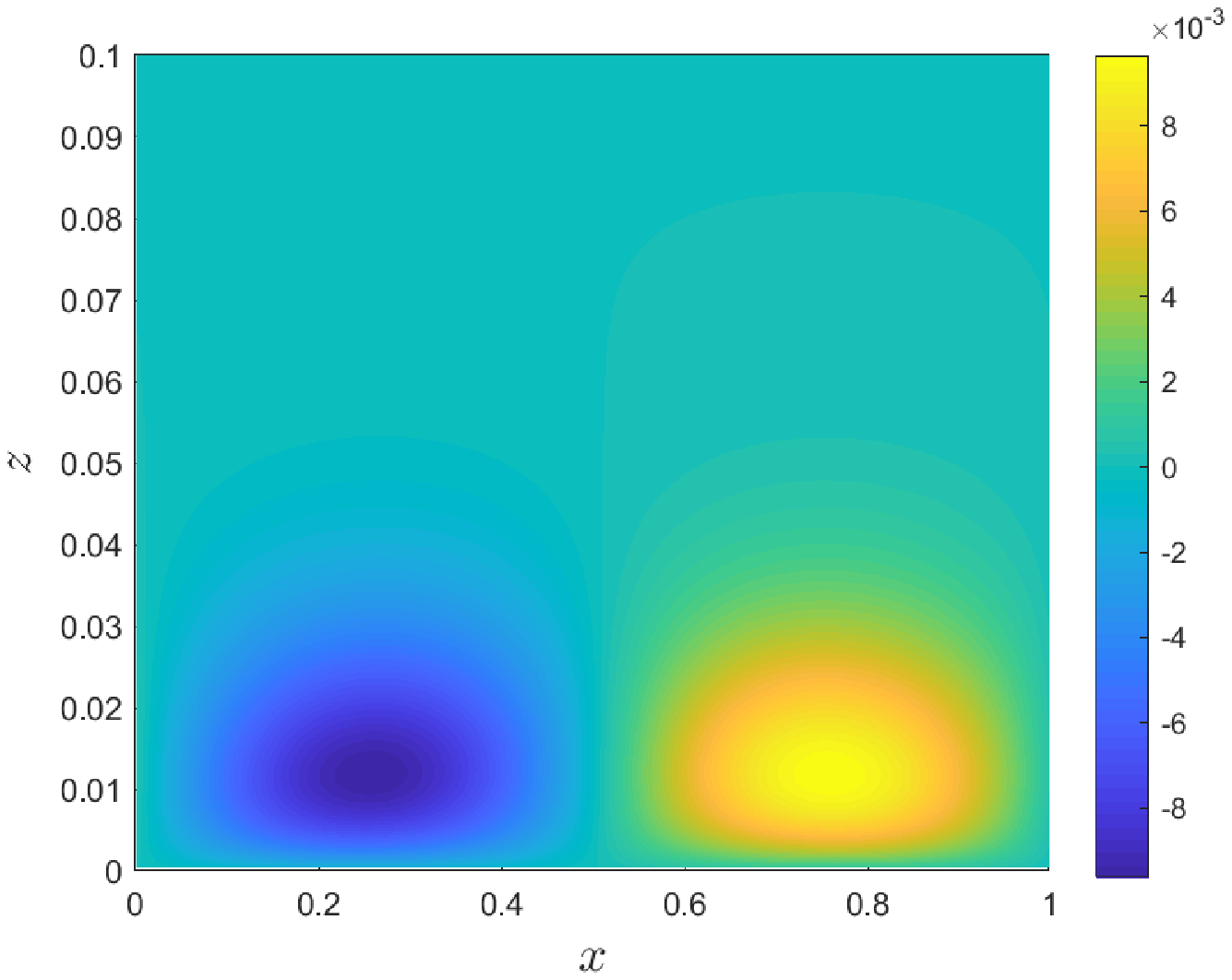}}\hfill
  \subfloat[$u_3$]
  {\includegraphics[width=0.25\textwidth, height=0.2\textwidth,
    clip]{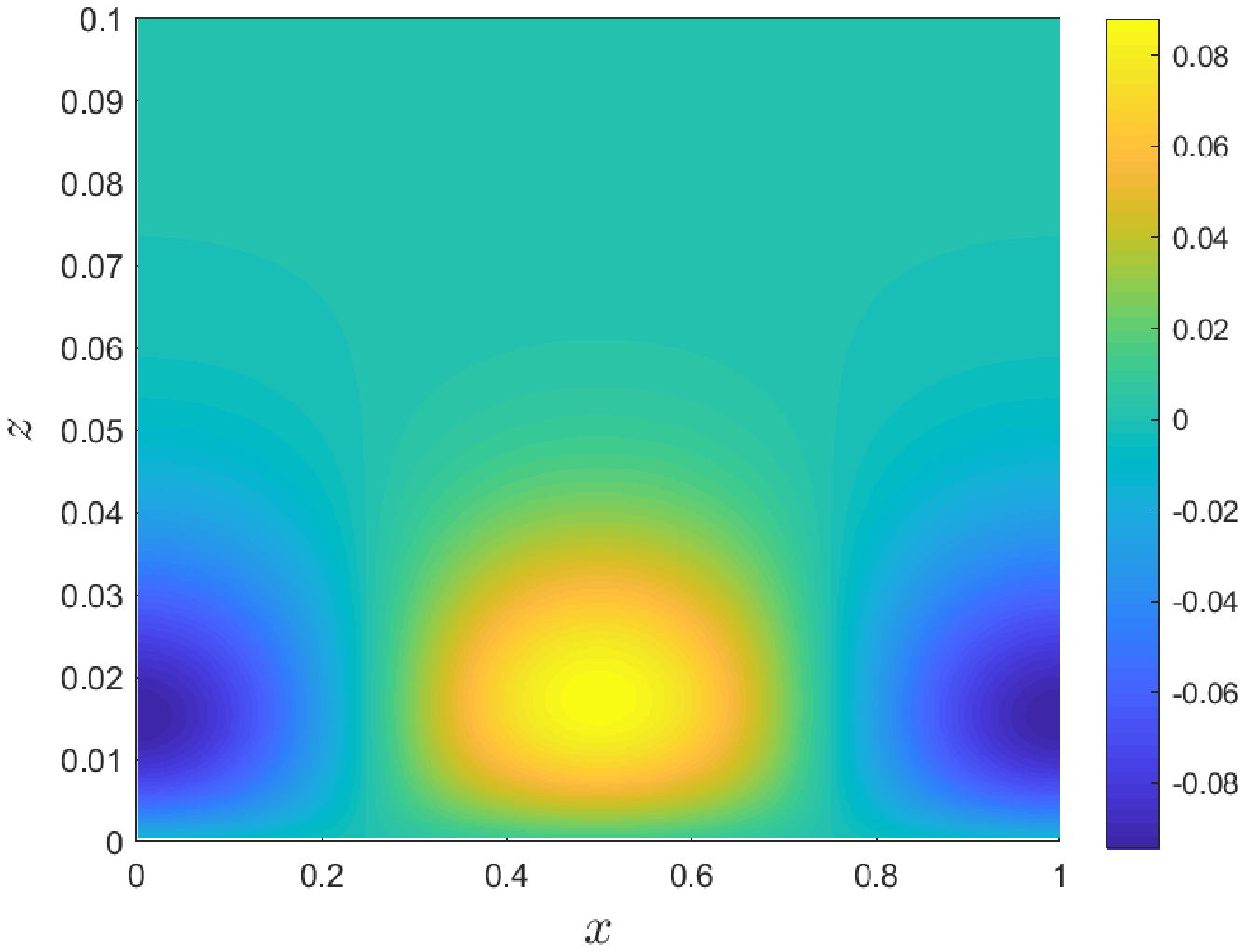}}\hfill
  \subfloat[$\mathcal{T}$]
  {\includegraphics[width=0.25\textwidth, height=0.2\textwidth,
    clip]{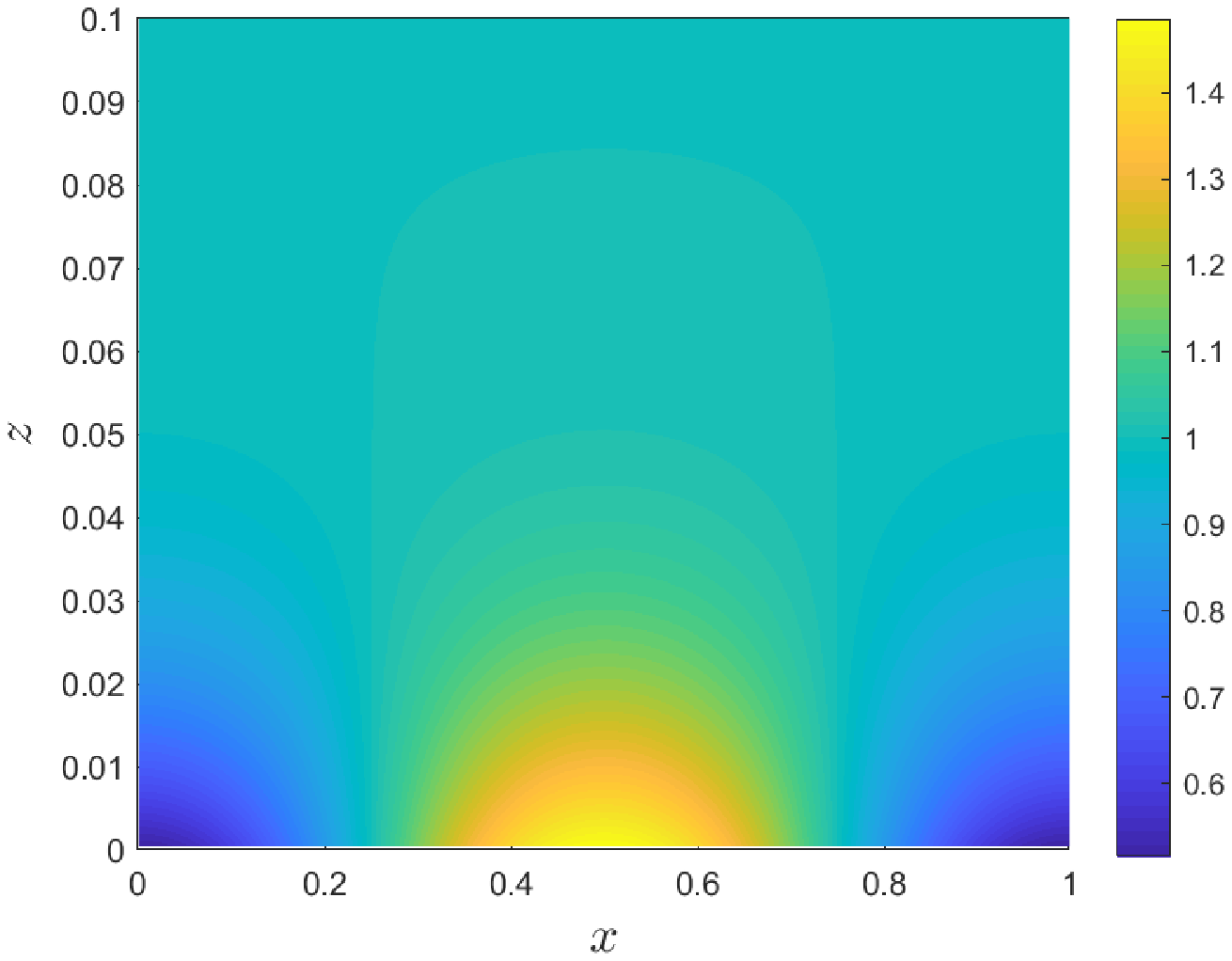}}\hfill
  \caption{Numerical solution with Prandtl number $\Pr = 2/3$ at
    $t=0.01$.}
\label{fig:test2_contour}
\end{figure}

\begin{figure}[!htb]
  \centering
  \subfloat[$\rho, x=0$]
  {\includegraphics[width=0.25\textwidth, height=0.2\textwidth,
    clip]{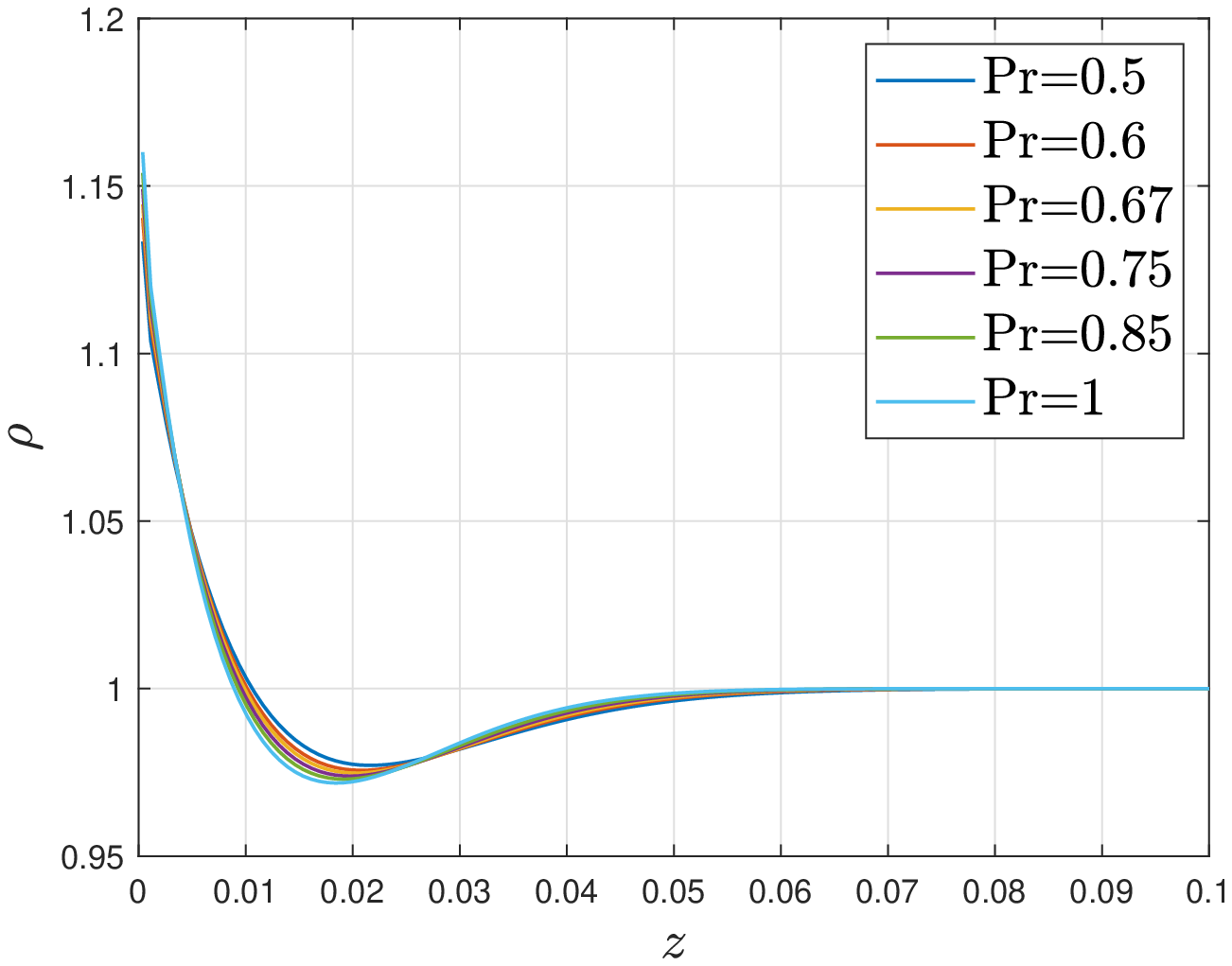}}\hfill
  \subfloat[$u_1, x=0$]
  {\includegraphics[width=0.25\textwidth, height=0.206\textwidth,
    clip]{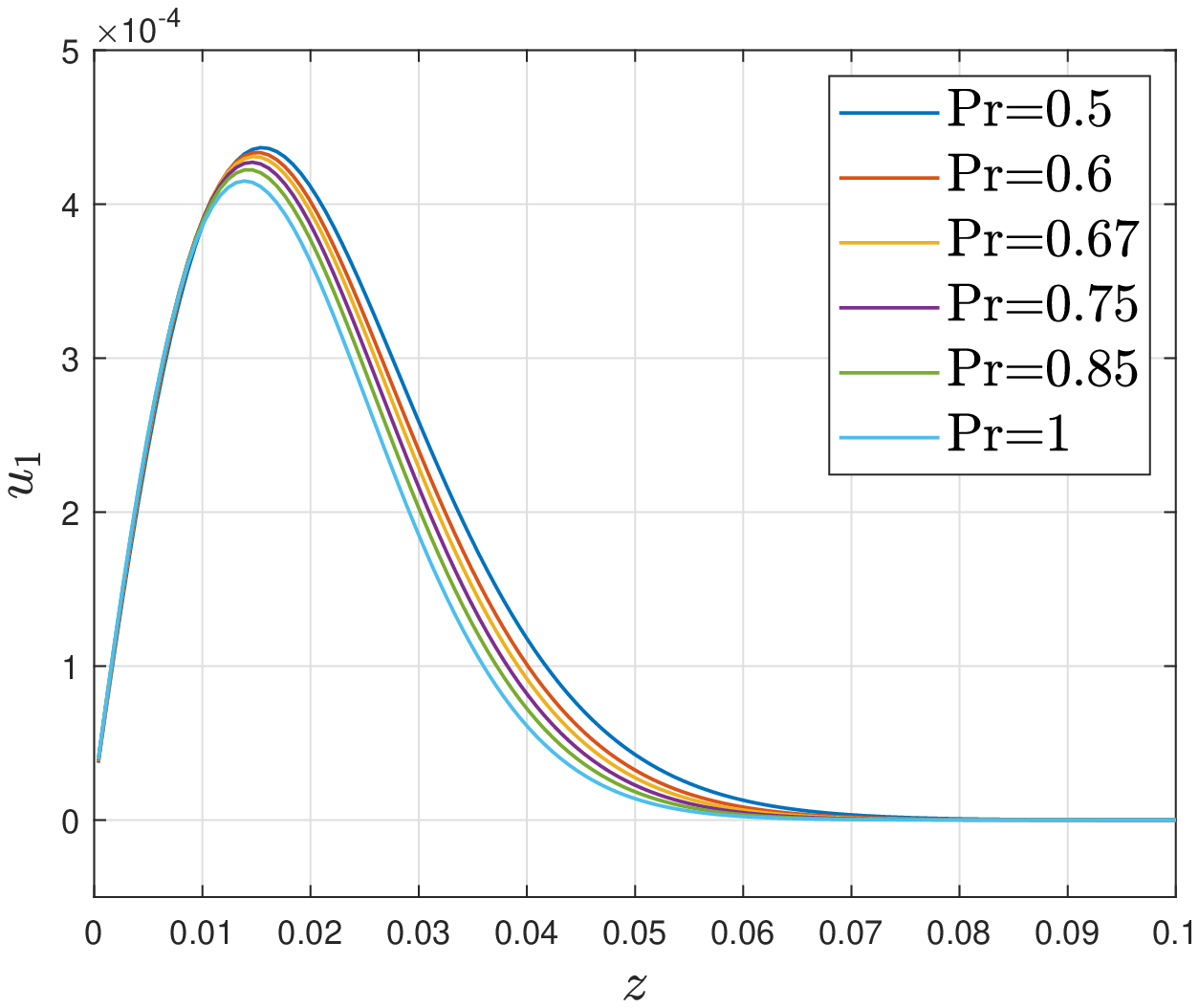}}\hfill
  \subfloat[$u_3,  x=0$]
  {\includegraphics[width=0.25\textwidth, height=0.2\textwidth,
    clip]{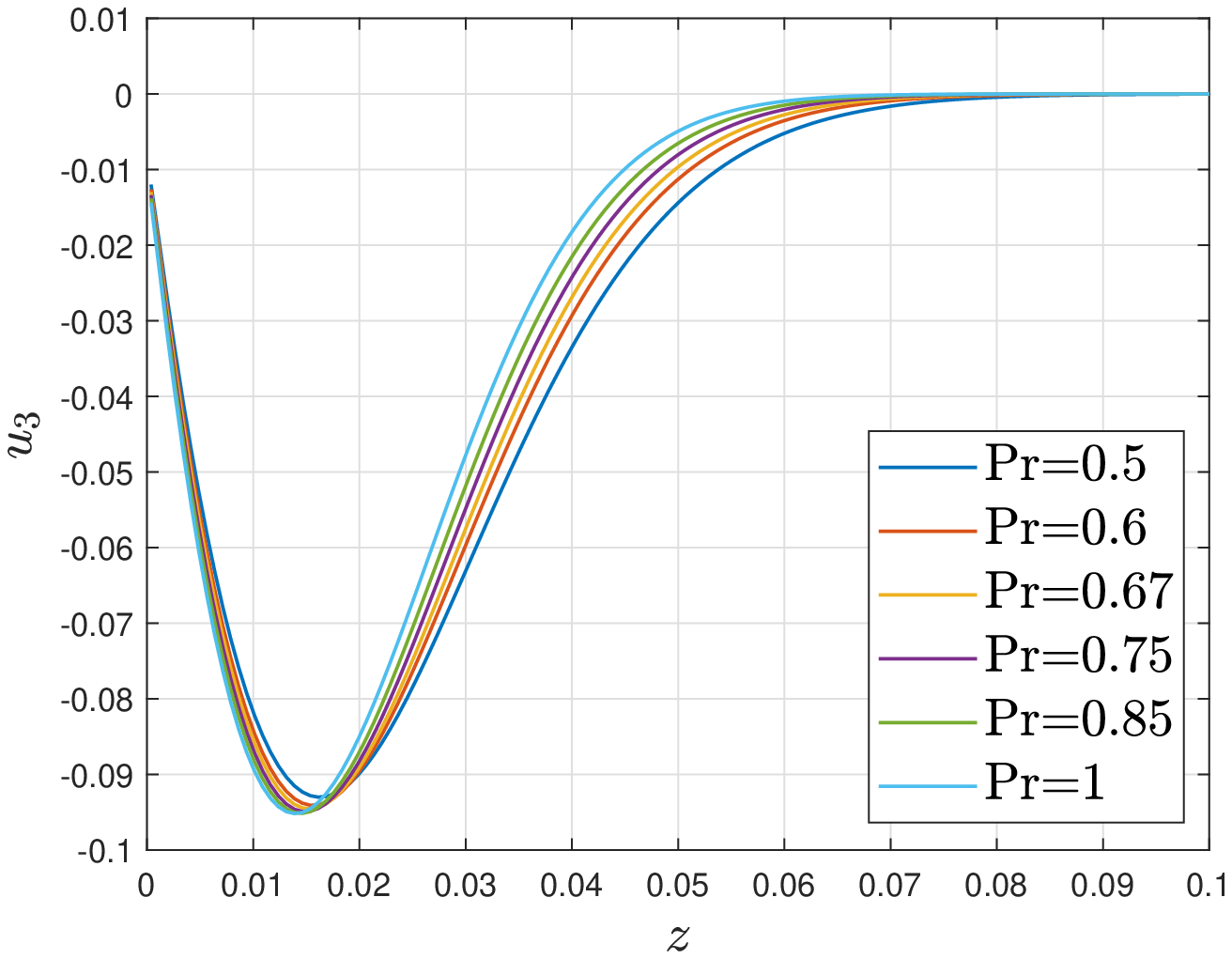}}\hfill
  \subfloat[$\mathcal{T}, x=0$]
  {\includegraphics[width=0.25\textwidth, height=0.197\textwidth,
    clip]{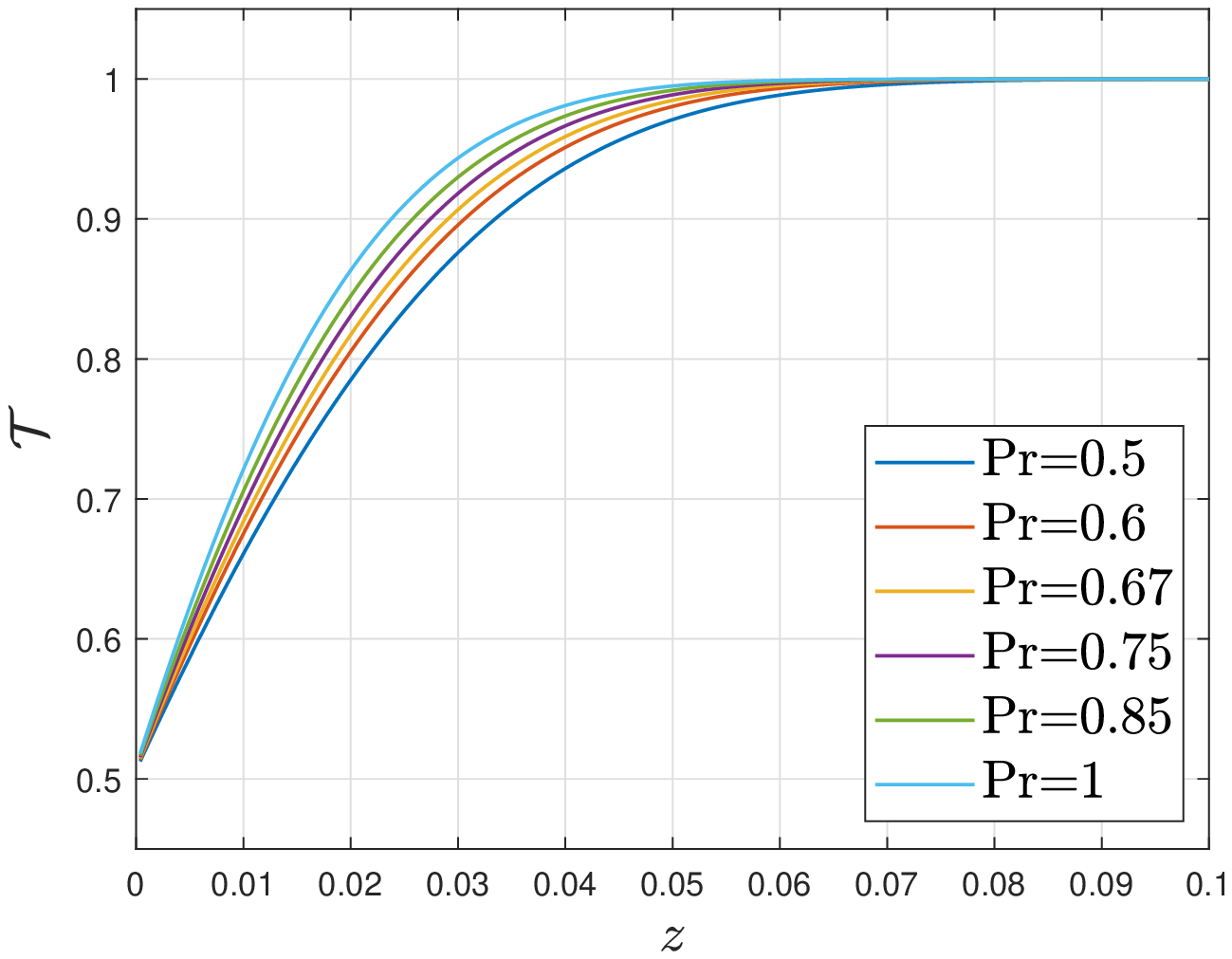}} \\
  \subfloat[$\rho, x=0.5$]
  {\includegraphics[width=0.25\textwidth, height=0.2\textwidth,
    clip]{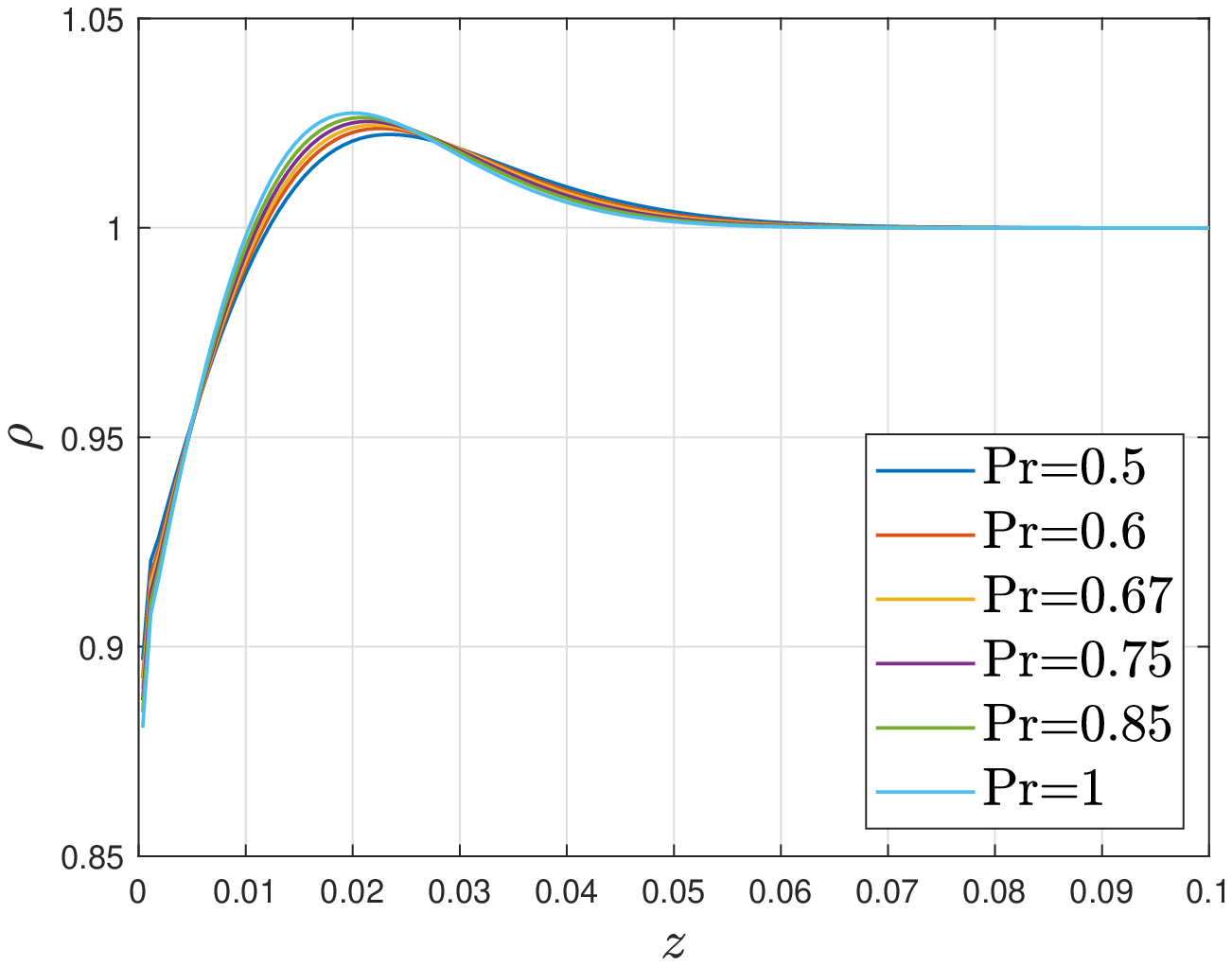}}\hfill
  \subfloat[$u_1, x=0.5$]
    {\includegraphics[width=0.25\textwidth, height=0.206\textwidth,
    clip]{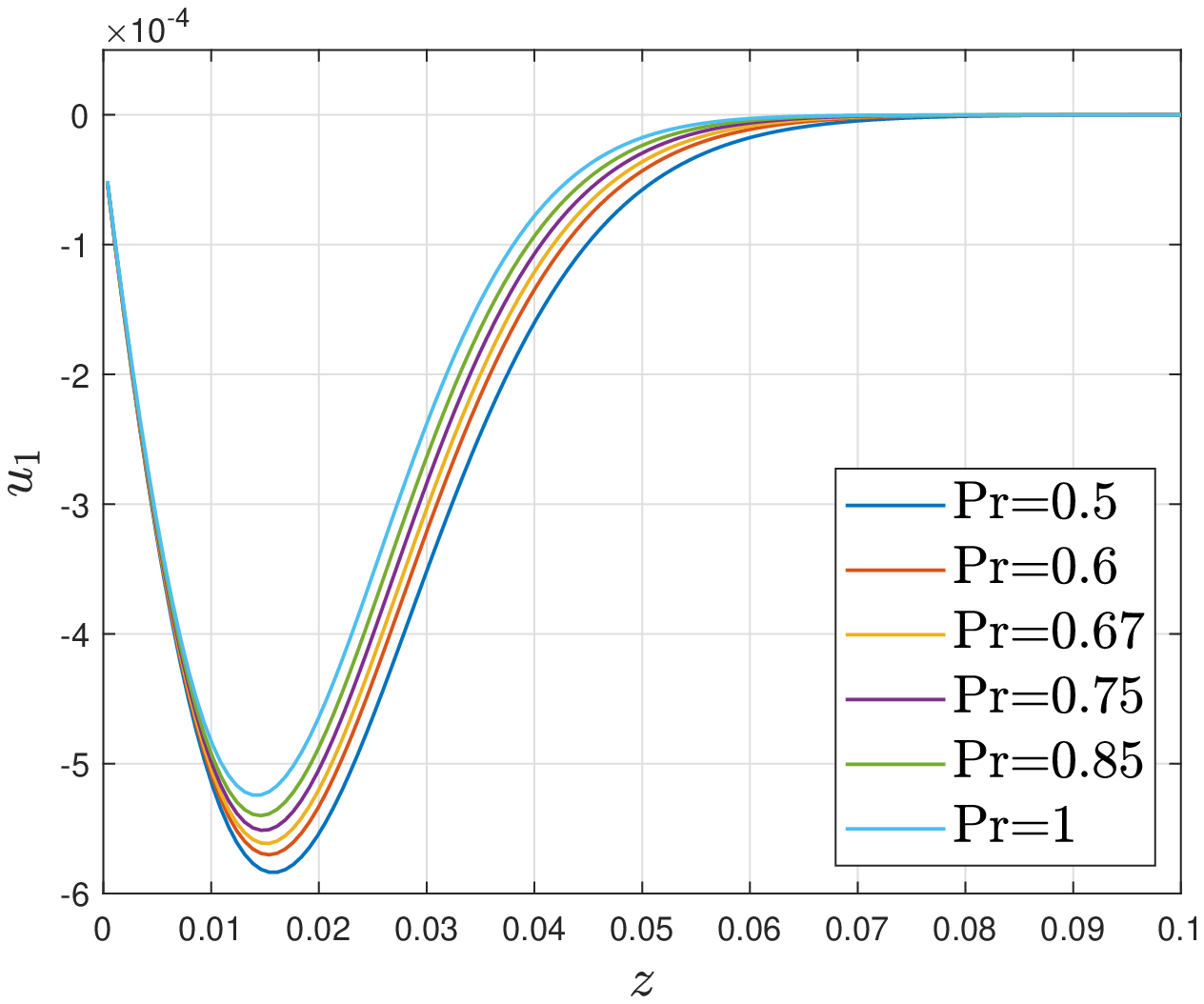}}\hfill
  \subfloat[$u_3,  x=0.5$]
  {\includegraphics[width=0.25\textwidth, height=0.197\textwidth,
    clip]{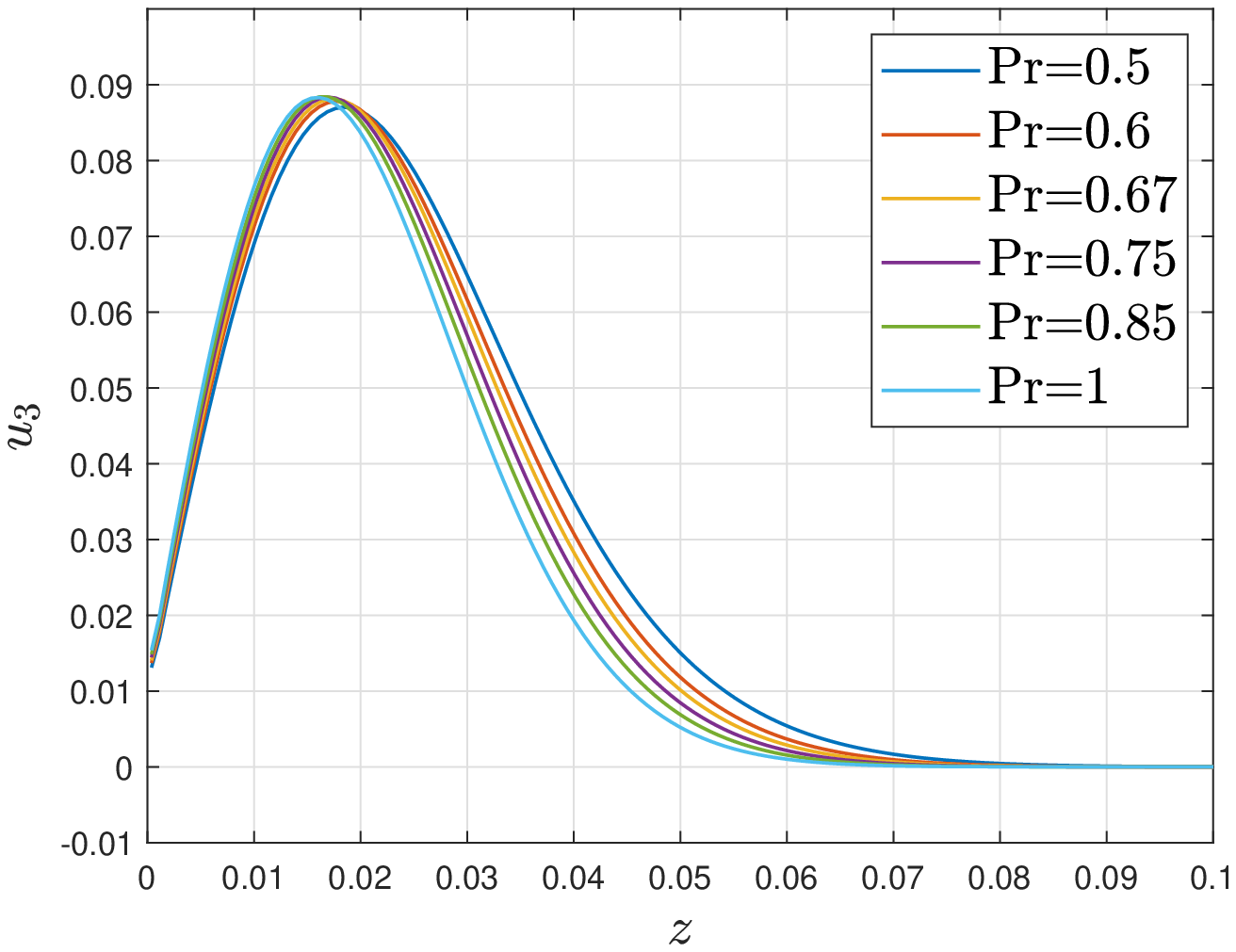}}\hfill
  \subfloat[$\mathcal{T}, x=0.5$]
  {\includegraphics[width=0.25\textwidth, height=0.197\textwidth,
    clip]{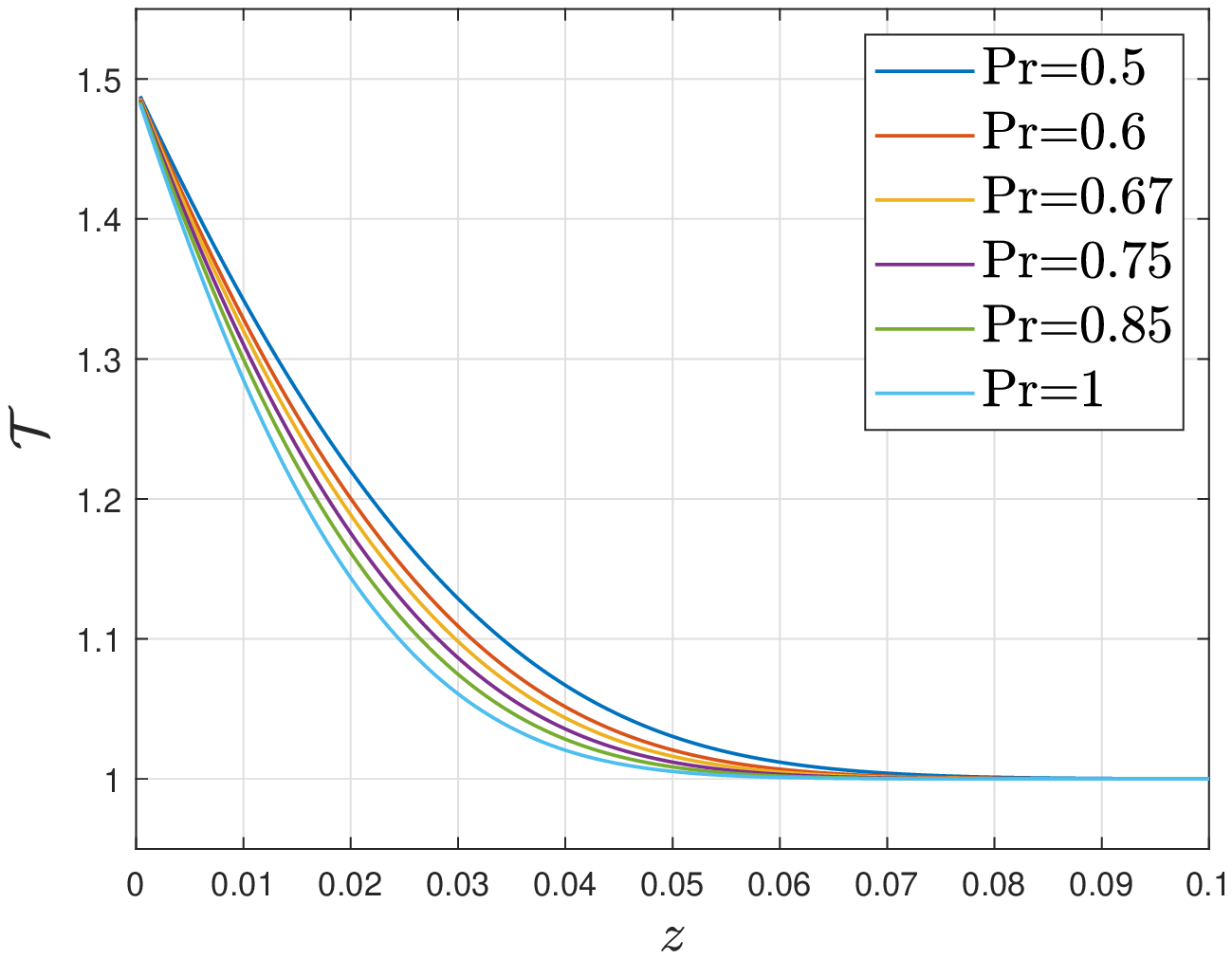}}
  \caption{Numerical solutions with different Prandtl numbers at
    $t=0.01$. The top column is at $x=0$, and the bottom column is at
    $x=0.5$.}
  \label{fig:test2_local}
\end{figure}

  In this test, we are focusing on the time-dependent problem, where a
  plasma set between two infinitely large plates is also studied, but
  one plate is at $z = 0$, and the other is at $z = +\infty$.  The
  periodic boundary is utilized in the $x$-direction. The setting of
  the problem is shown in Figure \ref{fig:test2}. A constant magnetic
  field of strength $B_y = 1$ acts in the direction of the $y$ axis.
  The density and velocity of the left wall are set as
  $\rho_{\rm left} = 1$ and $u_{\rm left} = 0$. The initial
  temperature is changing with $x$ as
  \begin{equation}
    \label{eq:temperature}
    \mT_{\rm left}  = 1 - 0.5\cos(2  \pi x).
  \end{equation}
  The initial
  condition of the particles is set as
  \begin{equation}
    \label{eq:initial_test2}
    \rho =1, \quad \bu = 0, \quad \mT = 1.
  \end{equation}
  The similar problem is studied in \cite{FILBET2018841, Sone1996} to
  show the ghost effect brought by the kinetic theory.  In this
  problem setting, there is also no gradient in $y$-direction, and we
  can derive
  \begin{equation}
    \label{eq:test2}
    \pd{\cdot}{y} = 0, \qquad u_2 = 0.
  \end{equation}
  The external electric field is added to balance the self-consistent
  electric field. Thus the system \eqref{eq:steady} is reduced into
  \begin{equation}
    \label{eq:steady2}
    \begin{aligned}
      & \pd{\rho}{t} + \pd{\rho u_1 }{x_1} + \pd{\rho u_3}{x_3} =  0, \\
      &\pd{u_1}{t} + u_1 \pd{u_1}{x_1}+ u_3 \pd{u_1}{x_3} - \frac{4\mu
      }{3\rho}\pd{^2 u_1}{x_1^2} - \frac{\mu}{\rho} \pd{^2u_1}{x_3^2}
      - \frac{\mu}{3\rho}\pd{^2 u_3}{x_1\partial x_3} + \pd{\mT}{x_1}
      + \frac{\mT}{\rho}\pd{\rho}{x_1}
      = -B u_3, \\
      &\pd{u_3}{t} + u_1\pd{u_3}{x_1} + u_3 \pd{u_3}{x_3} - \frac{4\mu
      }{3\rho}\pd{^2 u_3}{x_3^2} - \frac{\mu}{\rho}\pd{^2 u_3}{x_1^2}
      - \frac{\mu}{3\rho}\pd{^2 u_1}{x_1\partial x_3} + \pd{\mT}{x_3}
      + \frac{\mT}{\rho}\pd{\rho}{x_3}
      = B u_1, \\
      &\pd{\mT}{t} + u_1\pd{\mT}{x_1} + u_3 \pd{\mT}{x_3}+
      \sum_{i=1,3}\left[
        -\frac{8\mu}{9\rho}\left(\pd{u_i}{x_i}\right)^2
        -\frac{2\kappa}{3\rho} \pd{^2\mT}{x^2_i}
        + \frac{2}{3} \mT\pd{u_i}{x_i}\right] \\
      & \qquad \qquad -\frac{2\mu}{3\rho}
      \left(\pd{u_1}{x_3}+\pd{u_3}{x_1}\right)^2 + \frac{8\mu}{9\rho}
      \pd{u_1}{x_1} \pd{u_3}{x_3}= 0.
    \end{aligned}
  \end{equation}
  The numerical solutions at time $t = 0.01$ with different Prandtl
  numbers are studied. The upwind scheme is adopted to approximate the
  first-order derivatives with the central difference scheme to
  approximate the second-order derivatives in \eqref{eq:steady2}. In
  the numerical test, the grid size is $N=400$.  Due to the similar
  behavior of the solutions with different Prandtl numbers, Figure
  \ref{fig:test2_contour} shows the behavior of $\rho$, $u_1$, $u_3$
  and $\mT$ for Prandtl number $2/3$ with $\mu = 0.01$. We can find
  $u_3$ and $\mT$ are all changing periodically in the $x$-direction,
  which is consistent with the problem setting. To show the affection
  of Prandtl number, the local numerical solutions with different
  Prandtl numbers at different $x$ positions are plotted.  Figure
  \ref{fig:test2_local} illustrates the differences of the local
  solution $\rho$, $u_1$, $u_3$ and $\mT$ at position $x=0$ and $0.5$
  with different Prandtl numbers.

  We can find that for the time-dependent problem, even with a quite
  small time length, the numerical solutions vary greatly with
  different Prandtl numbers. This shows that Prandtl number will
  greatly affect the numerical solution and is an important factor to
  correctly describe the behavior of the plasma. The new reduced
  two-fluid model maintains the correct Prandtl number from the
  kinetic plasma model for any type of particles, and may be more
  capable to capture the behavior of the particles, which we will
  verify in future work.

\section{Conclusion Remarks}
\label{sec:conclusion}

We derived a new two-fluid model for the plasma using Maxwell
iteration based on the HME system. With the Shakhov collision model,
the new two-fluid model takes the correct Prandtl number in its
closure relations. Though it has the same degree of freedom as the
five-moment two-fluid model, this new reduced model is expected to
have improved performance for problems with anisotropic pressure
tensor and large heat flux, since anisotropic pressure tensor and heat
flux are expressed by the density, macroscopic velocity and
temperature instead of simply setting as zero. It is our future work
to investigate by numerical simulations the performance of the new
model in practical applications.

\section*{Acknowledgements}
We thank Prof. Yana Di at BNU-HKBU United International College for
the useful suggestions. The work of Ruo Li is partially supported by
the National Science Foundation of China (Grant No. 91630310) and
Science Challenge Project (No. TZ2016002). Yixiao Lu is partially
supported by the elite undergraduate training program of School of
Mathematical Sciences in Peking University. Yanli Wang is supported
by Science Challenge Project (No. TZ2016002) and National Natural
Science Foundation of China (Grant No. U1930402).


\bibliographystyle{plain}
\bibliography{../../article}
\end{document}